\documentclass[sigconf]{acmart}
\usepackage{enumitem}
\usepackage{setspace}
\usepackage{mdframed}
\usepackage{framed}
\usepackage{hyphenat}
\usepackage{algpseudocode}
\usepackage{amsmath,epsfig}
\usepackage[boxed,ruled,vlined,linesnumbered]{algorithm2e}
\usepackage{amsthm}
\usepackage[T1]{fontenc}
\usepackage{setspace}
\usepackage{enumitem}
\usepackage{ragged2e}
\usepackage{multirow}
\usepackage{hhline}
\usepackage{wrapfig}
\usepackage{amssymb}
\usepackage{caption}
\usepackage{subcaption}
\usepackage{graphicx}
\usepackage{amsmath,epsfig}
\usepackage{amsthm}

\setcopyright{rightsretained}
\usepackage{tikz}
\newcommand\encircle[1]{%
\tikz[baseline=(X.base)]
   \node (X) [draw, shape=circle, inner sep=-1.5pt, fill=black, text=white] {\strut #1};}

\newcommand{\B}{\vspace*{-\smallskipamount}}
\newcommand{\BB}{\vspace*{-\medskipamount}}
\newcommand{\BBB}{\vspace*{-\bigskipamount}}

\let\oldmathbf\mathbf
\makeatletter
\renewcommand{\mathbf}[1]{\oldmathbf{#1}\@ifnextchar_{\msubscript}{}}
\def\msubscript_#1{_{\!#1}}
\makeatother

\let\oldmathit\mathit
\makeatletter
\renewcommand{\mathit}[1]{\oldmathit{#1}\@ifnextchar_{\msubscript}{}}
\def\msubscript_#1{_{\!#1}}
\makeatother

\copyrightyear{2020}
\acmYear{2020}
\setcopyright{acmcopyright}
\acmConference[CODASPY '20]{Proceedings of the Tenth ACM Conference on Data and Application Security and Privacy}{2020}{New Orleans, LA, USA}
\acmBooktitle{Proceedings of the Tenth ACM Conference on Data and Application Security and Privacy (CODASPY '20), 2020, New Orleans, LA, USA}
\acmPrice{15.00}
\acmDOI{10.1145/3374664.3375737}
\acmISBN{978-1-4503-7107-0/20/03}

\begin{document}
\title{\textsc{IoT Expunge}: Implementing Verifiable Retention of IoT Data}\titlenote{\textbf{This paper has been accepted in 10th ACM Conference on Data and Application Security and Privacy (CODASPY), 2020.} \\ This material is based on research sponsored by DARPA under agreement number FA8750-16-2-0021. The U.S. Government is authorized to reproduce and distribute reprints for Governmental purposes notwithstanding any copyright notation thereon. The views and conclusions contained herein are those of the authors and should not be interpreted as necessarily representing the official policies or endorsements, either expressed or implied, of DARPA or the U.S. Government. This work is partially supported by NSF grants 1527536 and 1545071.}

\author{Nisha Panwar,$^{1,2}$ Shantanu Sharma,$^2$ Peeyush Gupta,$^2$ Dhrubajyoti Ghosh,$^2$ Sharad Mehrotra,$^2$ and Nalini Venkatasubramanian$^2$}
\affiliation{\institution{$^1$Augusta University, USA. $^2$University of California, Irvine, USA.}}

\renewcommand{\shortauthors}{Panwar et al.}
\renewcommand{\shorttitle}{\textsc{IoT Expunge}: Implementing Verifiable Retention of IoT Data}

\begin{abstract}
The growing deployment of Internet of Things (IoT) systems aims to ease the daily life of end-users by providing several value-added services. However, IoT systems may capture and store sensitive, personal data about individuals in the cloud, thereby jeopardizing user-privacy. Emerging legislation, such as California's CalOPPA and GDPR in Europe, support strong privacy laws to protect an individual's data in the cloud. One such law relates to strict enforcement of data retention policies. This paper proposes a framework, entitled \textsc{IoT Expunge} that allows sensor data providers to store the data in cloud platforms that will ensure enforcement of retention policies. Additionally, the cloud provider produces verifiable proofs of its adherence to the retention policies. Experimental results on a real-world smart building testbed show that \textsc{IoT Expunge} imposes minimal overheads to the user to verify the data against data retention policies.
\end{abstract}

\ccsdesc[500]{Security and privacy~Security protocols}
\ccsdesc[300]{Security and privacy~Mobile and wireless security}
\ccsdesc[300]{Security and privacy~Domain-specific security and privacy architectures}
\ccsdesc[100]{Security and privacy~Social aspects of security and privacy}

\keywords{Internet of Things; smart building; user privacy; data deletion; verification.}

\maketitle

\section{Introduction}
\label{sec:introduction}
The emerging Internet of Things (IoT) systems use sensors to create a digital representation of the state of the physical environments, individuals immersed in it, and their interactions with the physical space, as well as, with each other. Such a dynamic state representation provides a variety of value-added services to end-users and makes the existing processes (such as temperature control and knowing people locations) more efficient. While data captured by sensors is useful for service provisioning, it has significant privacy implications. Several studies~\cite{DBLP:journals/popets/ApthorpeHRNF19,DBLP:conf/edbt/Bertino16,DBLP:journals/jnca/SunCRSLYL17} have recently highlighted how sensor data can lead to unexpected inferences about individuals and their behavior. Regulations, such as General Data Protection Regulation (GDPR)~\cite{gdpr_ori}, California Online Privacy Protection Act (CalOPPA)~\cite{caloppa}, and California Consumer Privacy Act (CCPA)~\cite{ccpa}, have imposed several requirements on the organizations in which they can retain their user data. For instance, GDPR emphasizes data minimization, both in terms of the volume of the data stored of an individual and the duration of retainment. It states that personal data can only be kept for no longer than it is necessary for the purposes for which it is being processed. Making IoT systems compliant to such legislation, thus, poses an important challenge of ensuring that the underlying infrastructure implements data retention policies.

In this paper, we consider data retention for the use-case where sensor data providers outsource data to the cloud and provide access to the data to a variety of service providers that use the sensor data to provide different services. Examples of such a use-case scenario can be a cellular provider outsourcing customers' connectivity data to cell towers (from which their approximate location can be determined) to the cloud and providing such information to service providers (via the cloud) that build location-based services~\cite{DBLP:journals/jnca/SunCRSLYL17} based on such data. Another example could be mapping services that collect location data of individuals (\textit{e}.\textit{g}., via GPS on their mobile devices) and outsource the collected data to other service providers (\textit{e}.\textit{g}., location-based advertisers). One concrete context driving our solution is a university-based WiFi system, managed by the University Office of Information Technology (OIT), that collects and outsources users' connectivity information in order to allow
researchers to build smart space services (see \S\ref{sec:Experimental Evaluation} for the details of our live test-bed, called TIPPERS~\cite{DBLP:conf/percom/MehrotraKVR166}).

The data retention policy for such scenarios requires the cloud to delete/expunge the sensor data after a predefined period of time. For instance, policy for data captured by indoor surveillance cameras at our university is 4-days (to account for 3-day long weekends during which the university is closed), and the policy for WiFi connectivity data (that is often used to track missing/stolen phones by the police) is set in collaboration with the university police department. \textsc{IoT Expunge} framework, proposed in this paper, provides a mechanism for the cloud to produce a proof of deletion, thereby providing a verifiable implementation of the data retention policy. \textsc{IoT Expunge} enables any third party (whether it be the sensor data provider or the end-user whose data is captured by the sensors) to verify the correct implementation of the retention policy, without the use of a centralized trusted party.

\textsc{IoT Expunge} is not only applicable in the university IoT application settings (or similar IoT settings) as we discussed above, it can also be applied to other use-cases that require us to keep the data against retention policies in a verifiable manner. For example, a \emph{vehicle rental system} may use the verifiable data deletion mechanism provided by \textsc{IoT Expunge}. Particularly, rented vehicles contain an Event Data Recorder (EDR) that captures information about the itinerary or driving patterns of the drivers. However, a vehicle might be rented by different drivers at different points of time. Therefore, the parts of EDR data might belong to different drivers, who rented the vehicle. However, vehicle rental system requires that all such data related to a driver must be deleted as soon as the driver returns the vehicle, and mechanisms to empower the driver should verify the deletion would significantly enhance the security and trust of the user.


In this paper, we focus on building a verifiable data retention model for storing IoT data at the cloud. This problem deals with three sub-problems: \emph{timestamp generation} to allocate cryptographically verifiable timestamp to sensor records/readings (to verify them later); \emph{data state transition} to delete the data against the data retention policies; and \emph{attestation} to verify the state of sensor data against the data retention policies.

\medskip
\noindent\textbf{Contributions.} In this paper, we provide:
\begin{itemize}[noitemsep,nolistsep,leftmargin=0.01in]
  \item A framework to outsource sensor data to the cloud (\S\ref{sec:IoT Expunge_dataflow}), thereby service providers can develop applications using data, while users can verify the state of data against pre-notified data retention policies.
  \item A mechanism for allocating cryptographically verifiable timestamp (\S\ref{subsec:Control Phase}) to sensor records based on one-way accumulators~\cite{firs}.
  \item A verifiable data deletion/expunge protocol based on memory-hard functions~\cite{DBLP:conf/crypto/DworkN92,DBLP:conf/crypto/DworkNW05,DBLP:journals/toit/AbadiBMW05} (\S\ref{subsec:State Transition Phase} and \S\ref{subsec:Attestation Phase}), which do not exploit any trusted-party to execute verification.
  \item Performance evaluation (\S\ref{sec:Experimental Evaluation}) of \textsc{IoT Expunge} on university live WiFi data collected over 12 months.
\end{itemize}

\noindent\textbf{Outline of the paper.}
\S\ref{sec:Preliminary} provides an overview of entities involved in \textsc{IoT Expunge}, the threat model, the security goals. \S\ref{sec:Cryptographic Primitives} provides an overview of cryptographic building blocks, namely one-way accumulators and memory-hard functions, which will be used in our protocol development. We begin describing \textsc{IoT Expunge} by restricting it for the case when only a single service provider accesses the encrypted sensor data from the cloud (\S\ref{sec:IoT Expunge_dataflow} and \S\ref{sec:complete protocol}).

\noindent
\textbf{Full version.} In the full version~\cite{TR2020} of this paper, we show how such a model can be extended to support multiple service providers with different data retention policies.

\section{Preliminaries}
\label{sec:Preliminary}
This section provides the entities involved in \textsc{IoT Expunge}, the threat model, and security properties.

\subsection{Entities}
\label{subsec:Entities}
\begin{figure}[!t]
\centering
\includegraphics[scale=0.26]{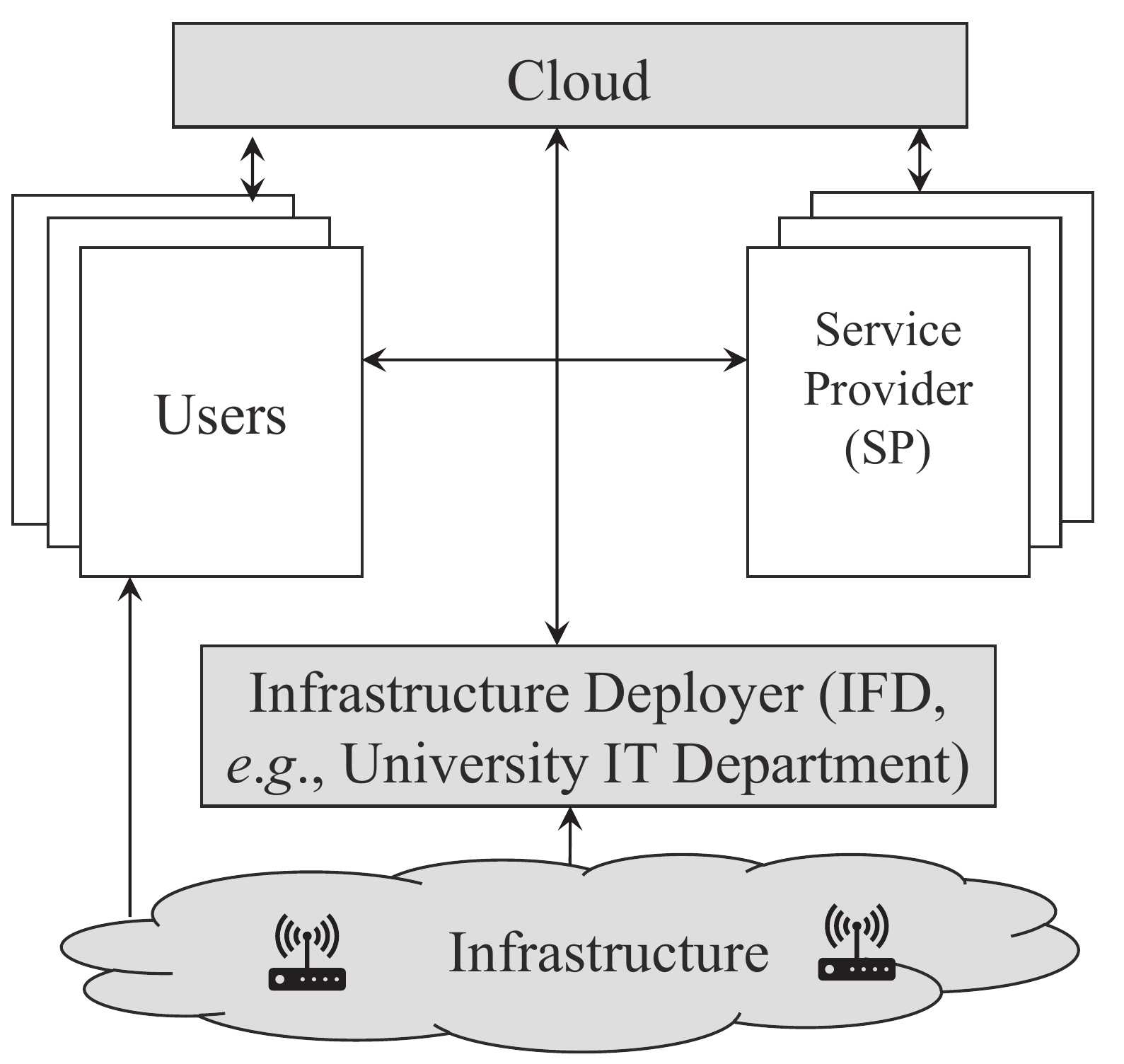}
\BB
\caption{Entities in \textsc{IoT Expunge}.}
\label{fig:entity}
\BB\BBB
\end{figure}

Our model has the following entities: sensor data provider (SDP), the public cloud, service providers (SP), and users; see Figure~\ref{fig:entity}.

\medskip
\noindent\textbf{Sensor Data Provider (SDP).} SDP (which is the university OIT department in our use-case; see \S\ref{sec:introduction}) deploys and owns a network of $p$ sensors devices (denoted by $s_1, s_2, \ldots, s_p$), which capture information related to users in a space.
In providing services, the sensors capture data related to the user; for instance, a WiFi access-point, captures the user-device-id (\textit{e}.\textit{g}., MAC address), say $d_i$, when it gets connected to the access-point, say at time $t_k$, and it produces a sensor record denoted by $\langle d_i, t_k, \mathit{load}_k\rangle$, where $\mathit{load}_k$ may contain sensor device id or any other payload information. By the SDP, sensor records are allocated timestamps that are discretized into epochs using which data retention policies are specified (as will be described in \S\ref{subsec:Data Retention Policy}). Before sending the sensor data to the cloud, the SDP encrypts it non-deterministically~\cite{DBLP:journals/jcss/GoldwasserM84}.

\medskip
\noindent\textbf{The public cloud.} The public cloud stores the encrypted sensor data received from the SDP. The cloud allows access to encrypted sensor data to service providers (SPs). Only those SPs that already have negotiated with the sensor provider about the sensor data usage, are given the data by the cloud. The data at the cloud remains accessible to the SPs, until the data expiration time. After this time, the data is deleted. (We will consider additional access control policies, wherein different SPs can have differentiated accesses to data, as an extension in \S6 of the full version~\cite{TR2020}).

\medskip
\noindent\textbf{Service Providers (SPs).} The SPs access encrypted sensor data based on their agreement with the SDP from the cloud. In our running example of university WiFi connectivity data, the SP corresponds to the TIPPERS system that accesses WiFi connectivity data to provide location-based services (see \S\ref{subsec:Testbed Description} for details of the TIPPERS system). Data provided to an SP is non-deterministically encrypted, and the SP cannot decrypt the data. However, the SP contains a secure enclave~\cite{sgx} (which works as a trusted agent of the SDP) using which the SP can provision services over encrypted data.\footnote{{\scriptsize Since secure enclave is a trusted agent of SDP, it can decrypt and compute over encrypted data. There are challenges in computing using enclaves due to side-channel attacks, \textit{e}.\textit{g}., cache-line, branch shadow, page-fault attacks~\cite{DBLP:conf/ccs/WangCPZWBTG17}, but since the focus of this paper is on implementing data retention policies, we do not address those challenges in this paper.}} The SP may request encrypted data from the cloud prior to the data being deleted.

\medskip
\noindent\textbf{Users.} Let $u_1,u_2,\ldots,u_{m^{\prime}}$ be the users who carry $m$ devices (denoted by $d_1,d_2,\ldots,d_m$), where $m^{\prime}\leq m$. Using these devices, users enjoy services provided by SDP, as well as, by SP. We define a term \emph{user-associated data} as follows: let $\langle d_i, t_k, \mathit{load}_k\rangle$ be a sensor reading, where $d_i$ be the $i^{\mathit{th}}$ device-id owned by a user $u_i$. We call the sensor reading $\langle d_i, t_k, \mathit{load}_k\rangle$ as user-associated data with the user $u_i$.

\subsection{Data Retention Policy}
\label{subsec:Data Retention Policy}
A data retention policy specifies the duration of time for which a cloud can store the sensor data. We model the data retention policy using the concept of epochs. In particular, timestamps are discretized into epochs using which data retention policies are specified. An epoch, denoted by $T_i$, is identified as a range of time $[\mathit{T_i.\mathit{bt}}, T_i.\mathit{et}]$ based on its begin time ($\mathit{bt}$) and end time $\mathit{et}$, and all sensor readings during that time period are said to belong to that epoch. There are no gaps between two consecutive epochs, \textit{i}.\textit{e}., the end time of the previous epoch is the same as the begin time of the next epoch. Thus, we can identify each epoch by its beginning time $T_i.\mathit{bt}$. For simplicity, we will assume that each epoch is of an equal duration, \textit{i}.\textit{e}., $\forall i, j, (T_i.\mathit{et} - T_i.\mathit{bt} = T_j.\mathit{et} - T_j.\mathit{bt})$. We refer to the duration of an epoch $T_i$ as: $\Delta = T_i.\mathit{et} - T_j.\mathit{bt}$. At any given time $t$, we refer to the epoch to which $t$ belongs as $Epoch(t)$, \textit{i}.\textit{e}., $\mathit{Epoch}(t).\mathit{bt} \leq t \leq \mathit{Epoch}(t).\mathit{et}$.

We model a data retention policy as a pair $\langle \mathcal{P}_{\mathit{del}}, \mathcal{P}_{\mathit{ver}}\rangle$, where $\mathcal{P}_{\mathit{del}}$ corresponds to the number of epochs after which the data must be deleted, and $\mathcal{P}_{\mathit{ver}}$ corresponds to the number of epochs until which the cloud must support mechanisms to verify the deletion (the value of $\mathcal{P}_{\mathit{ver}}$ can be set of infinity). After $\mathcal{P}_{\mathit{ver}}$ epochs, the deleted data cannot be verified, since it might be removed from the storage.\footnote{{\scriptsize At present, such policies are used by many cloud providers, \textit{e}.\textit{g}., Dropbox, that completely remove data from the storage media, if a person does not access the data for more than one year.}} More formally, assume a sensor data generated at time $t$, which belongs to an epoch, denoted by $\mathit{Epoch}(t)$. Such a sensor data must be deleted by the cloud at the beginning of an epoch whose begin time is $\mathit{Epoch}(t).\mathit{et} + \mathcal{P}_{\mathit{del}} \times \Delta$. Furthermore, the cloud must maintain enough information to enable a third party to verify deletion, until the beginning of epoch whose begin time is identified by $\mathit{Epoch}(t).\mathit{et} + \mathcal{P}_{\mathit{ver}} \times \Delta$.

\medskip
\noindent\textbf{Note: Data States.} Based on the data retention policies, the data can be in one of the two states: \emph{accessible} and \emph{irrecoverable}.
Prior to deletion, the sensor data is said to be in an \emph{accessible} state at the cloud. A sensor data that has been properly deleted by the cloud is said to be in an \emph{irrecoverable} state and cannot be accessed by the SP. The data that is in irrecoverable state cannot be converted into an accessible state. The data in both states can be verified by the user against data retention policy, prior to $\mathcal{P}_{\mathit{ver}}$.

\begin{figure}[!t]
\BBB
\centering
\includegraphics[scale=0.34]{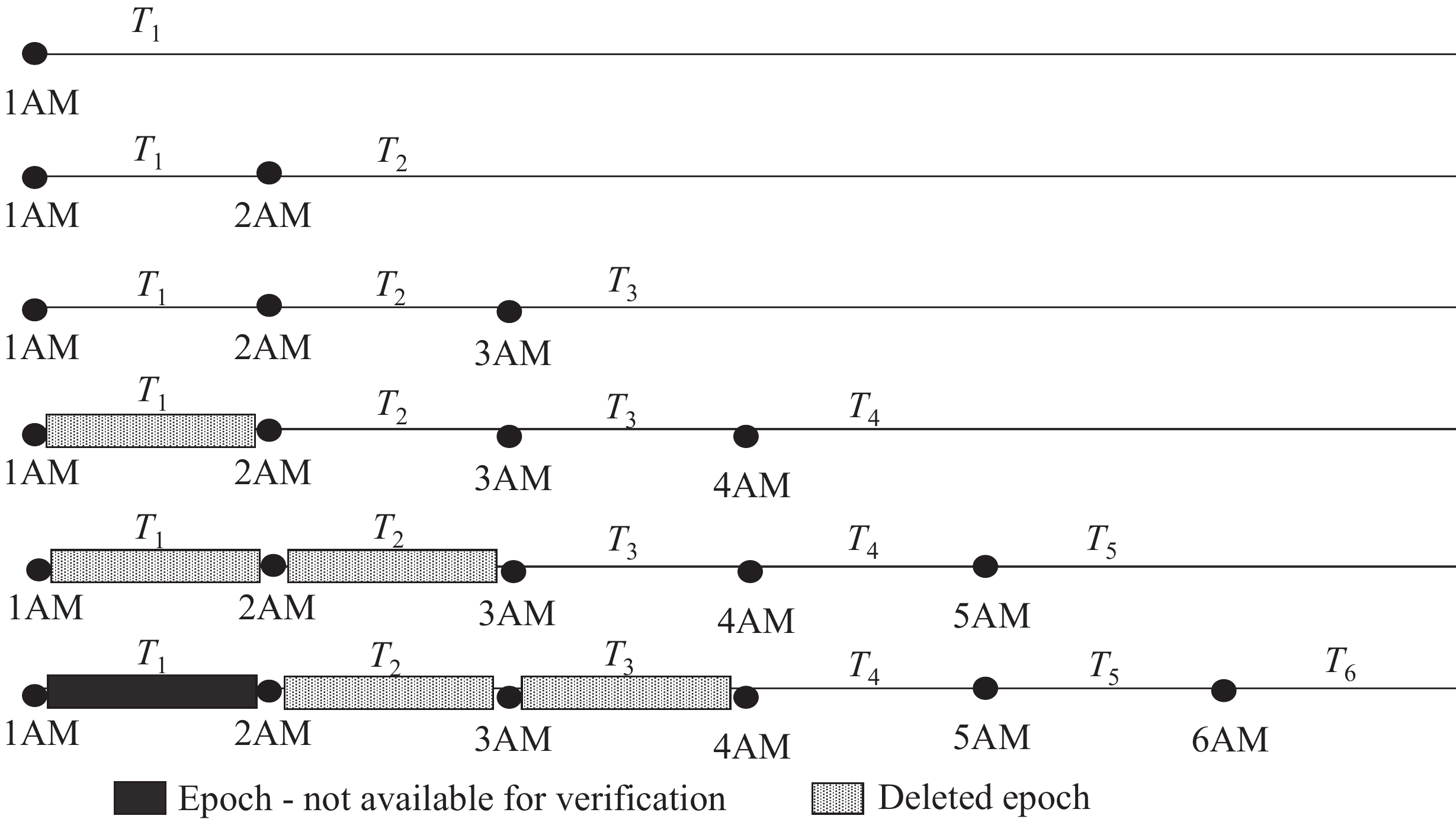}
\BBB\BBB\BB
\caption{An example illustrating the data retention policy.}
\label{fig:An example illustrating the data retention policy}
\BB
\end{figure}

\noindent\textbf{Example: Data retention policy.} Figure~\ref{fig:An example illustrating the data retention policy} shows an example of data arrival, epoch creation, and data deletion. In this example, we assume $\mathcal{P}_{\mathit{del}}=2$, $\mathcal{P}_{\mathit{ver}}=4$, and $\Delta=1$. In Figure~\ref{fig:An example illustrating the data retention policy}, each dot shows either the start or end of an epoch.

Note that as the data arrives in the epoch $T_4$, the data of the epoch $T_1$ is deleted (\textit{i}.\textit{e}., the state of data of the epoch $T_1$ becomes irrecoverable), since $\mathit{Epoch}(t).\mathit{et} + \mathcal{P}_{\mathit{del}} \times \Delta= 2+2\times 1=4$, where $t$ is a time value that belongs to the epoch $T_1$. Similarly, as the data arrives in the epoch $T_5$, the data of the epoch $T_2$ is deleted, since $\mathit{Epoch}(t).\mathit{et} + \mathcal{P}_{\mathit{del}} \times \Delta= 3+2\times 1=5$, where $t$ is a time value that belongs to the epoch $T_2$. However, the state of data of the epochs $T_1$ and $T_2$ still can be verified against the data retention policy.

When the data arrives in the epoch $T_6$, the data of the epoch $T_1$ may not be available for verification, since $\mathit{Epoch}(t).\mathit{et} + \mathcal{P}_{\mathit{ver}} \times \Delta= 2+4\times 1=6$, where $t$ is any time value belongs to the epoch $T_1$. However, it is important to mention that the data that belongs to the epoch $T_1$ cannot be converted into the accessible state.

\subsection{Threat Model and Security Properties}
\label{subsec:Threat Model}
\noindent
\textbf{Threat Model.} We assume that SDP and sensor devices are trusted and are secure. That is, sensors cannot be spoofed and, furthermore, malicious entities cannot launch an attack against the SDP to modify sensor data.\footnote{{\scriptsize We also assume a correct identification of sensors, before accepting sensor-generated data at SDP, and it ensures that no rogue sensor device can generate the data on behalf of an authentic sensor.}} We do not consider cyber-attacks that can exfiltrate data from the SDP, since defending against such attacks is outside the scope of this paper. 
Also, we assume that except for the SDP and the secure enclave at the SP, no other entity can learn the secret key to decrypt the sensor data.

The public cloud is assumed not to be malicious, \textit{i}.\textit{e}., it correctly executes the tasks requested to it by the SP (\textit{e}.\textit{g}., request for data) and by the SDP (\textit{e}.\textit{g}., store/delete data based on the data retention policy).\footnote{{\scriptsize We assume the existence of an authentication protocol between the SDP and the cloud, so that the SDP sends the data to a designated cloud. Further, an authentication protocol exist between the cloud and the SPs, so that the cloud forwards encrypted sensor data to only desired SPs.}} However, the cloud is not trusted to delete data correctly and must support verification for deletion. Such a cloud-based model is known as \emph{trust-but-verify} and widely considered in many cryptographic
algorithms~\cite{DBLP:journals/ieeesp/LinsGSS16,DBLP:journals/tsc/DongLW16,DBLP:conf/mobisys/WilsonWCBLW17}.
The trust-but-verify model is motivated by situations (\textit{e}.\textit{g}., GDPR), where the cloud provider wishes to protect itself against spurious litigation about violating data retention policy by providing verifiable proof of deletion.

We assume that an SP may behave maliciously. As mentioned before that the SP utilizes sensor data to provide services to the user, but SP may \emph{mimic} the user behavior by asking queries to learn the encrypted sensor data. A user may also behave maliciously and wishes to learn about the encrypted data during data state verification against the data retention policies. Note that a user may also learn about the data by asking queries to the SP about other users; however, we do not focus on such issues, since our focus is on the verifying data against the data retention policies.

\medskip
\noindent
\textbf{Security Properties.} In the above-mentioned threat model, an adversary wishes to show that the cloud is not behaving against the data retention policy. Hence, we need to develop a verification mechanism that can (\textit{i}) prove the cloud keeps sensor data according to the data retention policy, and (\textit{ii}) prevent any information leakage about the sensor data during the verification process. Thus, we need to maintain the following properties in our system:

\noindent\textbf{Privacy-Preserving Verification.} As the state of the sensor data changes at the cloud, the cloud should produce a proof to show it adheres to the data retention policy. The verification/attestation mechanism must prove that the cloud is executing the desired (deletion) task, against the data retention policy. However, the verification process must not reveal any information about other users to preserve their privacy.

\smallskip
\noindent\textbf{Minimality.} The verification process must be communication efficient, in terms of not providing the entire sensor data to the verifier (to attest the data state). The verification process must request the minimal amount of the data from the cloud, that is sufficient to verify the data state for the requested time period.

\smallskip
\noindent\textbf{Immutability.} We need to maintain immutability of queries arriving from the user to verify that the SP is not executing the queries by mimicking the user. Note that if the SP can alter the query log, it can execute any query, while no entity can detect such a behavior of the SP. Thus, having an immutable query log provides a way to detect malicious behavior of the SP.

\smallskip\noindent
\emph{Aside.}~\S\ref{subsec:Control Phase},~\S\ref{subsec:State Transition Phase}, and \S\ref{subsec:Attestation Phase} develop a protocol that ensures privacy-preserving verification property while maintaining minimality property.~\S\ref{subsec:Query Log Phase} provides a protocol to produce immutable query logs.

\subsection{Scoping the Problem}
\label{subsec:Scoping the Problem}
In general, implementing data retention policies on the cloud consists of two complementary tasks:
(\textit{i}) since cloud infrastructure may consist of several levels of caches, techniques need to be developed to track all the replicas of data (or data derived from the original data) and to expunge the data from all the replicas and/or caches, and
(\textit{ii}) techniques need to be designed to delete data from the storage media in such a way that the original data cannot be recovered from the deleted representation. Simply replacing data by a constant string (\textit{e}.\textit{g}., NULL string) or encrypting the data, as is commonly done today by cloud providers does not suffice, as shown in~\cite{gutmann1996secure,DBLP:conf/fast/PetersonBHSR05}.

For both the above-mentioned tasks, given the trust-but-verify model, the cloud will need to support mechanisms for verification. We scope the paper to address the above-mentioned second problem, wherein the cloud supports cryptographic protocols for verification of deletion from the storage media. Verifiable mechanisms to expunge data from caches and/or replicas potentially requires designing of verifiable data structures that maintain the links to all the copies of data, which is a significant independent problem in itself. In the remainder of the paper, we will assume that data on the cloud exists only on a single storage device, and our goal is to design methods to verify deletion of data from the storage, based on data retention policies provided by the SDP.



\section{Cryptographic Primitives}
\label{sec:Cryptographic Primitives}
Before describing \textsc{IoT Expunge} in detail, this section presents a brief overview of two existing cryptographic techniques, which we use in building \textsc{IoT Expunge}.

\noindent{\bf One-way Accumulators.} One-way accumulators were proposed by Benaloh and Mare
~\cite{firs} and are based on RSA assumption~\cite{DBLP:journals/cacm/RivestSA83}. We use the cryptographic RSA-based accumulators to construct a timestamping protocol that allocates cryptographically verifiable timestamp to each sensor reading. Here, we provide an overview of one-way accumulators that satisfy the quasi-commutative property. A quasi-commutative function $\textit{f}:X \times Y \rightarrow X$ can be defined as:

\centerline{$f(f(x,y_1),y_2)=f(f(x,y_2)y_1); \: \forall x\in X, \: \forall (y_1,y_2)\in Y$}
Also, the quasi-commutative property is satisfied, if the function $f$ is replaced by a one-way hash function $\mathcal{H}$. Let us assume that the hash function $\mathcal{H}$ is initialized with a seed value $x$ and recurrent values ($y_1, y_2, \ldots, y_n$), then the accumulated hash digest is:

\centerline{$z=\mathcal{H}(\mathcal{H}(\mathcal{H}(\ldots \mathcal{H}(\mathcal{H}(\mathcal{H}(x,y_1),y_2),y_3),\ldots,y_{n-2}),y_{n-1}),y_n)$}
The output $z$ will be identical, even when the values $y_1, y_2, \ldots, y_n$ are permuted, while all hash functions are identical. As an advantage, the quasi-commutative functions do not require any central authority during timestamp verification, in our context, (as well as, provide a space-efficient alternative to digital signatures). To see why a central authority is not required, while using quasi-commutative functions, consider an example, where $n$ values $y_1, y_2, \ldots, y_n$ come from $n$ different users, and those values generate a final accumulated hash digest $z$. Assume that a user $u_j$ is assigned a partially accumulated hash digest $z_j$ with all $y_i$, where $1\leq i\leq n$ and $i\neq j$. The user $u_j$ is holding the value $y_j$, can be verified by checking, if $z=\mathcal{H}(z_j,y_j)$. We consider the one-way accumulators based on RSA assumption~\cite{DBLP:journals/cacm/RivestSA83}. Consider that the RSA function $\mathcal{E}(x,y)=x^y \bmod \eta$ underlies the assumption that given $\mathcal{E}(x,y)$, $y$, and $\eta$, where $\mathcal{E}$ is an encryption function; $x$ cannot be computed in polynomial time. Since recovering $x$ from $y$ is at least as hard as integer factorization, it can appropriately be used as the one-way hash functions.

\medskip
\noindent{\bf Memory-hard Functions.} We propose to use memory-hard functions~\cite{DBLP:conf/crypto/DworkN92,DBLP:conf/crypto/DworkNW05,DBLP:journals/toit/AbadiBMW05,DBLP:conf/uss/BiryukovK16} to delete encrypted sensor data at the cloud, during state transition phase. The memory-hard functions\footnote{{Our approach is independent of any particular memory-hard function. In fact, any function that allows verification of deletion based on policies can be used in \textsc{IoT Expunge}.}} execute a series of computations on the input value, such that each computation step in the series is tied with the computation at the previous step. Thus, the way of computing the final answer shows the inability to compute the output using some locally stored intermediate values, rather than the initial value. The final answer to these memory-intensive functions is pre-computed and serves verification purposes. Specifically, in memory-hard functions, the verifier selects a pair $\langle d, a\rangle$, where $d$ denotes the difficulty level of the function and $a$ denotes the correct answer to the function. The prover must solve the function on the input by executing $d$ steps of an assigned computation and must generate a solution, which should match with $a$. Note that such a computation cannot be parallelized and, thus, achieves the verifiable time-space trade-off during the computation~\cite{DBLP:conf/crypto/DinurN17}. Memory-hard functions have been, also, used in different scenarios, such as verifying the number of replicas of the data by using shortcut-free functions~\cite{rezaa} and verifying the encrypted data at the cloud using hourglass functions~\cite{amy}.


\section{\textsc{IoT Expunge} --- Dataflow}
\label{sec:IoT Expunge_dataflow}
This section presents a brief overview of different phases involved in \textsc{IoT Expunge} and dataflow among different entities; see Figure~\ref{fig:Dataflow and computation in the protocol}:

\begin{figure}[!t]
\BB
	\begin{center}
	\includegraphics[scale=0.36]{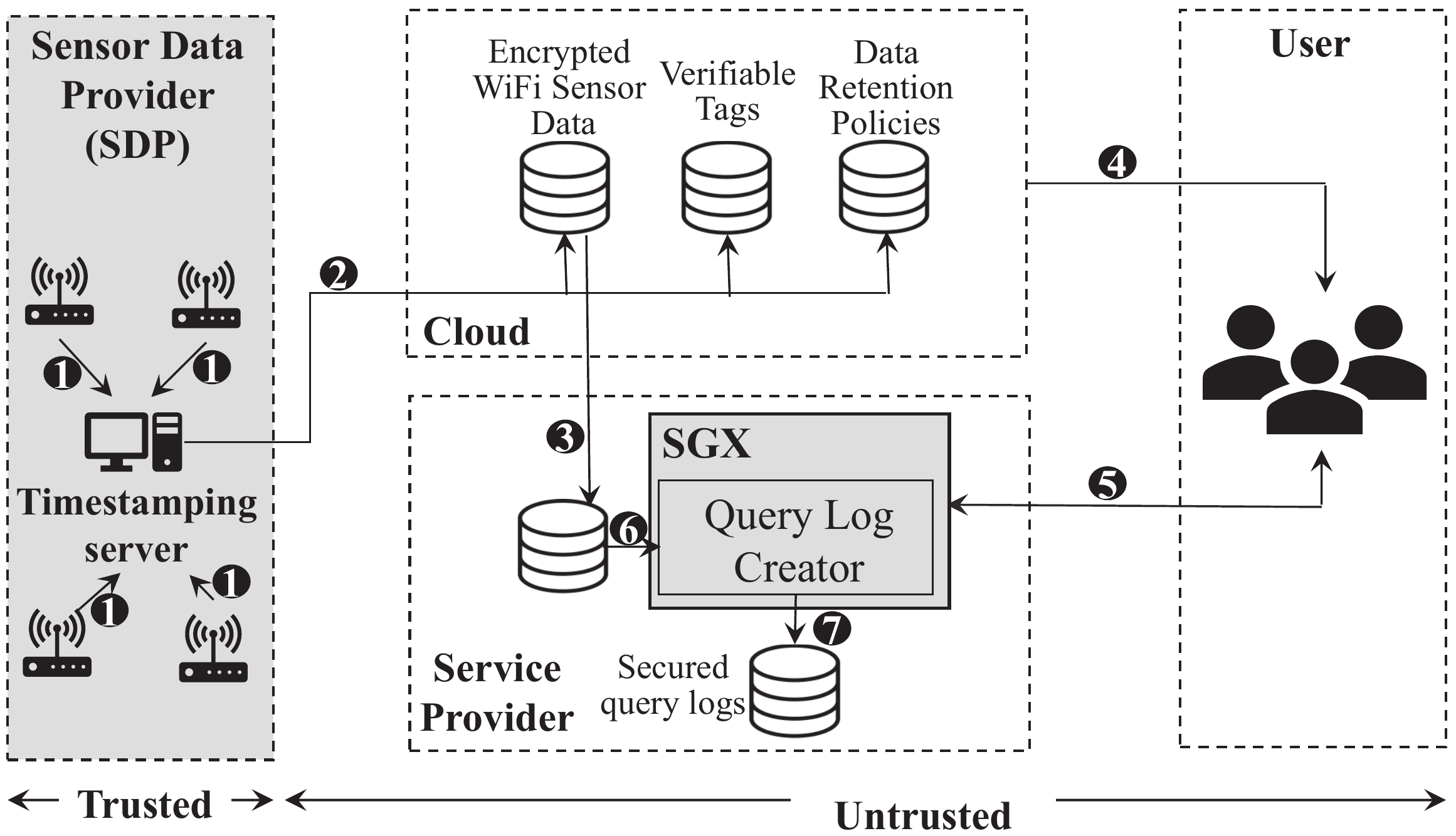}
	\end{center}
\BBB\B
	    \caption{Dataflow and computation in the protocol. Trusted parts are shown in shaded boxes.}
	\BBB
	\label{fig:Dataflow and computation in the protocol}
\end{figure}

\medskip
\noindent\textbf{Control phase: Dataflow from sensors to the SDP and the SDP to the cloud.} The SDP is equipped with a \emph{timestamping} server that collects all sensor readings/records (\encircle{1}). The timestamping server partitions the timeline into multiple epochs having a range of time $[\mathit{T_i.bt}, \mathit{T_i.et}]$, where $\mathit{bt}$ and $\mathit{et}$ denote the begin time and end time, respectively, of the epoch $T_i$. The timestamping server allocates the same epoch-id ($T_i$) to all sensor readings that belong to the epoch $T_i$ having the time range $[\mathit{T_i.bt}, \mathit{T_i.et}]$. The timestamping server, also, appends a \emph{cryptographic timestamp}, denoted by $\mathcal{CT}$. Further, the SDP generates \emph{verifiable tags}. Cryptographic time and verifiable tags are used to attest the data state (accessible and irrecoverable). The SDP outsources following to the cloud (\encircle{2}): (\textit{i}) encrypted sensor data, (\textit{ii}) verifiable tags, (\textit{iii}) a list of SPs who can access the encrypted data, and (\textit{iv}) data retention policies. The SDP may keep sensor data in cleartext or in encrypted form. This phase is entitled \emph{control phase} (see \S\ref{subsec:Control Phase} for details).


\medskip
\noindent\textbf{State transition phase: Dataflow from the cloud to the SP.} The cloud stores the data received from the SDP. As mentioned previously that we will, first, build \textsc{IoT Expunge} for only \emph{a single SP} (in \S\ref{sec:complete protocol}), in which case the data can reside in \emph{accessible} and \emph{irrecoverable} states.
The sensor data in accessible state can be accessed by all SPs. The sensor data in irrecoverable state cannot be accessed by all SPs. The cloud converts the data state against data retention policies and generates verifiable proofs to show that it adheres to the data retention policies. This phase is entitled \emph{state transition phase} (see \S\ref{subsec:State Transition Phase} for details). Further, the cloud sends encrypted data to all the designated SPs (\encircle{3}).

\begin{figure*}[!t]
\centering
\includegraphics[scale=0.66]{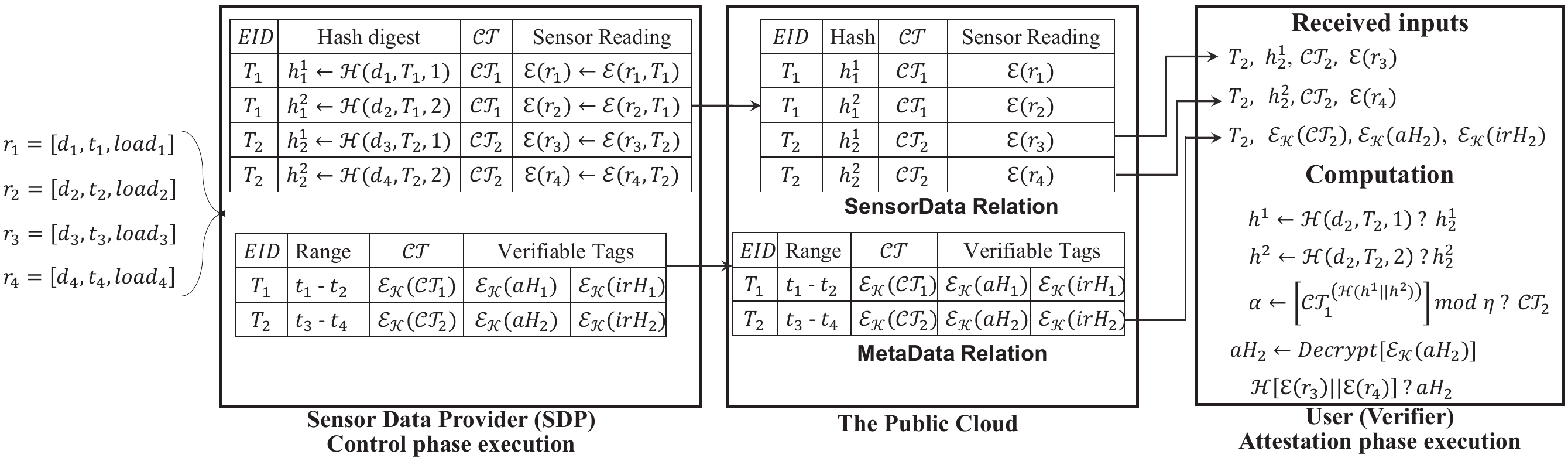}
\BBB\BBB
\caption{\textsc{IoT Expunge} protocol execution.}
\label{fig:protocol execution}
A user wishes to verify his/her data at time $t_3$. Notations:
$\mathit{EID}$: Epoch-id,
Range: Begin/end time of an epoch,
$\mathcal{CT}_1\leftarrow x^{\mathcal{H}(h^1_1||h^2_1)}$, $\mathcal{CT}_2\leftarrow \mathcal{CT}^{\mathcal{H}(h^1_2||h^2_2)}$, and ?: Comparing values.
\BBB
\end{figure*}

\medskip
\noindent\textbf{Attestation phase: Dataflow from the cloud to the users.}
Users wish to verify the state of their encrypted sensor data at the cloud against data retention policies. Thus, the cloud sends the verifiable tags corresponding to the desired epoch (\encircle{4}), using which the user verifies her/his data, without involving in
a heavy computation at their end. This phase is entitled attestation
phase (see \S\ref{subsec:Attestation Phase} for details). Further, the SDP can also verify the data state.

\medskip
\noindent\textbf{Query logging phase: Dataflow from the user to the SP.} The SP stores encrypted sensor data (\encircle{3}), received from the cloud. For building services, the SP has the secure enclave (Intel Software Guard eXtension, SGX~\cite{sgx}) that works as a trusted agent of the SDP. The secure enclave receives the digitally signed user queries (\encircle{5}) and provides answers after decrypting the data inside the enclave and processing the sensor data (\encircle{6}). On receiving a query, the enclave stores the query and the identity of the user with its digital signature on the disk in a secure and tamper-proof manner, to prevent the SP to execute a query by impersonating a real user (\encircle{7}). This phase is entitled \emph{query logging phase} (see \S\ref{subsec:Query Log Phase} for details).


\section{\textsc{IoT Expunge} --- The Protocols}
\label{sec:complete protocol}
This section provides details of \textsc{IoT Expunge} for the case where only one SP exists, and the state of encrypted sensor data at the cloud changes from accessible to irrecoverable. 
Figure~\ref{fig:protocol execution} shows a complete execution of \textsc{IoT Expunge} protocol over four sensor readings.


\medskip
\noindent\textbf{Preliminary phases: Key distribution, user-device registration, and data retention policy broadcast:}
Before using \textsc{IoT Expunge}, there is a need of executing the following preliminary steps:

\medskip
\noindent\emph{\underline{Key distribution phase.}} We assume a key distribution phase that distributes public keys ($\mathit{PK}$) and private keys ($\mathit{PR}$). The trusted SDP (which is the university IT department in our setup of the TIPPERS system) generates/renews/revokes keys used by the secure (hardware) enclave (denoted by $\langle \mathit{PK}_E, \mathit{PR}_E\rangle$). The SDP uses $\mathit{PK}_E$ to encrypt sensor readings before sending them to the cloud.\footnote{{\scriptsize Following the existing frameworks~\cite{amaranth}, the sensor devices may itself generate an encrypted sensor data that can be decrypted by SDP for executing control phase. However, we do not consider such a model in this paper.}} $\mathit{PR}_E$ is used by the enclave to decrypt sensor readings.
The public key and private key of the SDP are denoted by $\mathit{PK}_{\mathit{SDP}}$ and $\mathit{PR}_{\mathit{SDP}}$, respectively. Further, the SDP shares an identical symmetric key with all users, say $\mathcal{K}$, which is used to securely encrypt the verifiable proof/tag for deletion process.

\medskip
\noindent\emph{\underline{User device registration.}} We also assume a registration process, thereby a user device registers itself to the SDP and the SP. For instance, in the radio-frequency identification (RFID) card system, users are identified by their RFID card, and the registration process consists of users providing the details of their cards and other identifiable information (\textit{e}.\textit{g}., email address or phone number). In the case of a WiFi network, users are identified by their mobile devices, and the registration process consists of users providing the MAC addresses of their devices and other identifiable information.

\medskip
\noindent\emph{\underline{Data retention policy broadcast.}} We assume that when SDP establishes a new data retention policy, it informs about it to all registered users using their provided email addresses or phone numbers, as well as, to the cloud. Also, SDP informs the hash function to the user and the cloud, which was used in the control phase by the SDP.

\subsection{Control Phase}
\label{subsec:Control Phase}
The control phase (see Algorithm~\ref{alg:The control phase execution}) is the first phase of \textsc{IoT Expunge}, where the SDP receives sensor records. The objective of this phase is to: (\textit{i}) partition the timeline into multiple epochs where each epoch consists of same duration of time\footnote{{\scriptsize For simplicity, we consider that the duration of each epoch is same.}} (\textsc{Stage} 1), (\textit{ii}) allocate an epoch-id and a cryptographic time to each sensor records belonging to the same epoch (\textsc{Stage} 2 and \textsc{Stage} 3), thereby the verifier can later verify the state of sensor readings in the dataset against the data retention policy;
(\textit{iii}) encrypt sensor data before outsourcing to the cloud (\textsc{Stage} 4). To achieve these objectives, the control phase contains four different stages, as follows:

\medskip
\noindent\textbf{\textsc{Stage} 1: Epoch creation.} The first stage finds an appropriate epoch duration. Recall that an epoch $T_i$ consists of a time range of $[\mathit{T_i.bt}, \mathit{T_i.et}]$, based on which sensor readings having the time in this range belong to the epoch $T_i$. Each epoch is allocated an epoch-id, which is the begin time of the epoch. In this paper, we denote an epoch-id by $T_i$, instead of $\mathit{T_i.bt}$ to simplify the notation.

The duration of an epoch impacts the following: the number of sensor records in the epoch, the communication cost between the cloud and a verifier, and the execution time of the attestation phase. Particularly, as the epoch duration increases, several sensor records are allocated to a single epoch (under the assumption that the arrival rate of sensor readings is uniform). Thus, in turn, as will be clear in \S\ref{subsec:Attestation Phase} and \S\ref{sec:Experimental Evaluation}, verifying a sensor record belonging to an epoch with a longer duration requires more (verification) time and communication cost (due to the data movement between the cloud and the verifier, during the attestation phase).



\textbf{Aside.} IoT sensor data may be generated at a different velocity at different time. For example, in the case of WiFi connectivity data arriving from an access-point associated with a building, several sensor readings are produced at fast speed in daytime, compared to WiFi data arriving from the same access-point in the nighttime. Our method can also deal with such a case, by creating epochs of different lengths of duration.


\medskip
\noindent\textbf{\textsc{Stage} 2: Time allocation (Lines~\ref{ln:allocate_epoch_time}-\ref{ln:crypto_time_for_non1} of Algorithm~\ref{alg:The control phase execution}).} All the sensor readings that belong to the same epoch are allocated a single cryptographic timestamp, which is generated using the one-way accumulator~\cite{firs}. In the following, we explain the steps in detail:

\smallskip
\noindent\textit{\underline{Step 1: Initialization.}} Initially, a seed value $x$ is generated using a pseudo-random number generating (PRG) function. Also, two large prime numbers $p$ and $q$ are generated as private values of the one-way accumulator, such that $\eta=p\times q$. Both the values $x$ and $\eta$ are public values.

\smallskip
\noindent\textit{\underline{Step 2: Cryptographic timestamp generation.}} Based on the cryptographic timestamp, we create a chain of timestamps using the one-way accumulator. Below, we explain the procedure for two successive epochs:

\noindent\textit{Cryptographic timestamp generation for the first epoch.} Consider that the first epoch has $n$ sensor readings that have allocated the epoch-id $T_1$. To generate the cryptographic time, say $\mathcal{CT}_1$, for the first epoch, we first compute hash digests,
for each sensor reading by executing a hash function, $\mathcal{H}$, over each device-id, the epoch-id, and a counter variable ($1$ to $n$), as follows (Line~\ref{ln:allocate_epoch_time} of Algorithm~\ref{alg:The control phase execution}):

\centerline{
$h^1_1 \leftarrow \mathcal{H}(d_i,T_1,1)$, \\
$h^2_1 \leftarrow \mathcal{H}(d_j,T_1,2)$, \\
$\ldots$, \\
$h^n_1 \leftarrow \mathcal{H}(d_k,T_1,n)$ \\
}
Where $h_i^j$ denotes the hash digest for the $j^{\mathit{th}}$ sensor reading of the $i^{\mathit{th}}$ epoch, and $d_i$, $d_j$, $d_k$ are user-device-ids associated with the first, second, and the $n^{\mathit{th}}$ sensor readings. Note that each hash digest will be different; hence, any adversarial entity cannot learn anything about the device behavior in the epoch. Then, we compute the cryptographic time $\mathcal{CT}_1$, which is allocated to all the $n$ sensor readings of the epoch, as
 follows (Line~\ref{ln:crypto_time_for_non1} of Algorithm~\ref{alg:The control phase execution}):

\centerline{
$\mathcal{CT}_1 \leftarrow [x^{\mathcal{H}(h^1_1||h^2_1||\ldots ||h^n_1)} ]\bmod \eta$}

\noindent\textit{Cryptographic timestamp generation for the second epoch.} Consider that the second epoch has $n^{\prime}$ sensor readings that have allocated the epoch-id, say $T_5$. To generate the cryptographic time, say $\mathcal{CT}_2$, for the second epoch, we compute hash digests, as we computed for the previous epoch:

\centerline{
$h^1_2 \leftarrow \mathcal{H}(d_i,T_5,1)$, \\
$h^2_2 \leftarrow \mathcal{H}(d_j,T_5,2)$, \\
$\ldots$, \\
$h^{n^{\prime}}_2 \leftarrow \mathcal{H}(d_k,T_5,n^{\prime})$
}

Then, we compute the cryptographic time, $\mathcal{CT}_2$, which is allocated to all the $n^{\prime}$ sensor readings, of the epoch, as follows (Line~\ref{ln:crypto_time_for_1}):

\centerline{$\mathcal{CT}_2 \leftarrow [\mathcal{CT}_1^{\mathcal{H}(h_2^1||h_2^2||\ldots ||h^{n^{\prime}}_2)} ]\bmod \eta$}

Note that here the cryptographic time $\mathcal{CT}_2$ is computed by using the cryptographic time $\mathcal{CT}_1$, which was computed for the first epoch. In a similar way, we can compute the cryptographic time for the third epoch and other epochs too (Line~\ref{ln:crypto_time_for_1}).

\medskip
\begin{mdframed}
Note that using the hash digest, verifiers can attest membership (absence/presence) of their sensor records in the epoch, and using the cryptographic timestamp, verifiers can attest the completeness of all hash digests produced during the epoch; (will be clear soon in \S\ref{subsec:Attestation Phase}).
\end{mdframed}
\medskip

\DontPrintSemicolon
\LinesNotNumbered
\begin{algorithm}[!t]
\textbf{Inputs:} Sensor reading $r_j=(d_j, t_j, \mathit{load}_j)$, where $j\in \{1, n\}$. Epoch: $T_i$. \\
Public values: $x$ and $\eta$. Hash function: $\mathcal{H}$. Public key of the enclave: $\mathit{PK}_E$. Encryption function: $\mathcal{E}$.

\textbf{Outputs:} Relations \texttt{SensorData} and \texttt{MetaData}

\nl{\bf Function $\mathit{Control\_phase}(r_j)$} \nllabel{ln:vrfy}
\Begin{
\nl \For{$j\in \{1,n\} \wedge j\in T_i$ \nllabel{ln:allocate_epoch_time}}{
\nl $h^j_i=\mathcal{H}(d_j, T_i, j)$ }
		
\nl  \For{$\forall T_i, i>0, $ \nllabel{ln:allocate_crpto_time}}{
			
\nl \lIf{$\neg T_1$}{
$\mathcal{CT}_i\leftarrow \mathcal{CT}^{\mathcal{H}(h^1_i||h^2_i||\ldots||h^n_i)}_{i-1} \bmod \eta$\nllabel{ln:crypto_time_for_1}}

\nl \lElse
{$\mathcal{CT}_1\leftarrow x^{\mathcal{H}(h^1_1||h^2_1||\ldots||h^n_1)}$}
\nllabel{ln:crypto_time_for_non1}}
		
\nl \For{$ j\in \{1,n\} \wedge j\in T_i$ \nllabel{ln:encrypt_readings}}{
	
\nl $\mathcal{E}_{\mathit{PK}_E}(r_j)\leftarrow \mathcal{E}_{\mathit{PK}_E}(d_j, t_j, \mathit{load}_j, T_j)$}
		
\nl $\mathit{aH}_i\leftarrow \mathcal{H}[\mathcal{E}_{\mathit{PK}_{E}}(r_j)] \: j\in \{1,n\}$ \nllabel{ln:verifiable_tag_ah_compute}
		
\nl $\mathit{irH}_i \leftarrow \mathcal{H}($Encrypted deleted rows after simulating deletion on $\mathcal{E}_{\mathit{PK}_E}(r_j))$: $j\in\{1,n\})$ \nllabel{ln:verifiable_tag_irh_compute}
		
\nl Outsource \texttt{SensorData} $\leftarrow \langle T_i, h^j_i, \mathcal{CT}_i, \mathcal{E}_{\mathit{PK}_{E}(r_j)}\rangle$ where $j\in\{1,n\}$ \nllabel{ln:outsource_sensordata}
		
\nl Outsource \texttt{MetaData} $\leftarrow \langle T_i,
\mathit{T_i.bt}, \mathit{T_i.et},
\mathcal{E}_{\mathcal{K}}(\mathcal{CT}_i), \mathcal{E}_{\mathcal{K}}(\mathit{aH}_i), \mathcal{E}_{\mathcal{K}}(\mathit{irH}_i)\rangle$ \nllabel{ln:outsource_metadata}
}
\caption{Control phase.}
\label{alg:The control phase execution}
\end{algorithm}
\setlength{\textfloatsep}{0pt}

\medskip
\noindent\textbf{\textsc{Stage} 3: Verifiable tags generation (Lines~\ref{ln:encrypt_readings}-\ref{ln:verifiable_tag_irh_compute} of Algorithm~\ref{alg:The control phase execution}).} As mentioned at the beginning of this section that the state of encrypted sensor data changes from accessible to irrecoverable. The SDP generates verifiable tags for each epoch, thereby a verifier (user/SDP) can verify the data state against data retention policies.

Below, we explain how the SDP produces verifiable tags for an epoch having $n$ sensor readings, denoted by $r_1=\{d_1,t_1,\mathit{load}_1, T_1\}$,
$r_2=\{d_2,t_2,\mathit{load}_2, T_1\}, \ldots,$
$r_n=\{d_n,t_n,\mathit{load}_n, T_1\}$, where $r_i$ denotes $i^{\mathit{th}}$ sensor reading; $d_i$, $t_i$, $\mathit{load}_i$ denote the $i^{\mathit{th}}$ user device, $i^{\mathit{th}}$ sensor time, and $i^{\mathit{th}}$ payload in the $i^{\mathit{th}}$ sensor reading; and $T_1$ denotes the epoch-id. Now, the verifiable tags for this epoch will be computed as follows:

\smallskip
\noindent\textit{\underline{Step 1: Encryption of the sensor records (Line~\ref{ln:encrypt_readings}).}} We first encrypt the sensor readings $r_1, r_2, \ldots r_n$ using the public key of the enclave, denote by $\mathcal{E}_{\mathit{PK}_E}(r_1), \mathcal{E}_{\mathit{PK}_E}(r_2), \ldots, \mathcal{E}_{\mathit{PK}_E}(r_n)$. For simplicity, from here on, we use the notation $\mathcal{E}(r_j)$ to denote $\mathcal{E}_{\mathit{PK}_E}(r_j)$, unless explicitly mentioned.

\smallskip
\noindent\textit{\underline{Step 2: Hash of encrypted data (Line~\ref{ln:verifiable_tag_ah_compute}).}} Now, we compute a hash function, $\mathcal{H}$, over the encrypted sensor readings:
$\mathit{aH}_1 \leftarrow \mathcal{H}[\mathcal{E}(r_1)|| \mathcal{E}(r_2)|| \ldots || \mathcal{E}(r_n)]$, where $\mathit{aH}_i$ denotes the hash digest for accessible state data of the epoch $i$.

\smallskip
\noindent\textit{\underline{Step 3: Simulate data deletion and compute hash digest (Line~\ref{ln:verifiable_tag_irh_compute}).}} Finally, SDP simulates the deletion process (described below) on the encrypted sensor readings $\mathcal{E}(r_1), \mathcal{E}(r_2), \ldots \mathcal{E}(r_n)$, computes a hash function on the output of the deletion process, and it results in a hash digest, denoted by $\mathit{irH}_i$, to indicate the hash digest for irrecoverable state data of the epoch $i$.

\medskip
\begin{mdframed}
Note that after knowing the membership of sensor data in an epoch, the verifier can attest the current state of all the sensor readings in the epoch using the verifiable tags.
\end{mdframed}
\medskip

\medskip
\noindent\textbf{\textsc{Stage} 4: Outsourcing data (Lines~\ref{ln:outsource_sensordata}-\ref{ln:outsource_metadata} of Algorithm~\ref{alg:The control phase execution}).} Now, the SDP has encrypted sensor readings of an epoch having the epoch-id $T_i$, cryptographic timestamp, and verifiable tags $\mathit{aH}_i$ and $\mathit{irH}_i$. All such information is outsourced to the public cloud in the form of two relations:
(\textit{i}) the first relation, called \texttt{SensorData}, contains
the epoch-id $T_i$,
hash digests for each sensor reading $h_i^y$ (where $y$ is the number sensor readings in the epoch),
cryptographic timestamp, and
sensor readings encrypted using the public key of the secure enclave ($\mathit{PK}_E$); and
(\textit{ii}) another relation, called \texttt{MetaData}
having the epoch-id, epoch begin/end time, cryptographic timestamp, and verifiable tags ($\mathit{aH}_i$ and $\mathit{irH}_i$). The SDP encrypts all fields of the \texttt{MetaData} relation using the key $\mathcal{K}$, except for epoch-id and epoch begin/end time.

\DontPrintSemicolon
\LinesNotNumbered
\begin{algorithm}[!t]
\textbf{Inputs:} $T_i$, $r_j \in T_i$, where $j\in \{1,n\} $,
Memory-hard function: \encircle{$\mathcal{H}$}

\textbf{Outputs:} Deleted sensor readings of $T_i$

\textbf{Variable initialization:}
$\mathit{Temp\_array}[]$,
$\mathit{iteration} \leftarrow \log n$,
$\mathit{stepSize} \leftarrow 1$, $\mathit{blockSize} \leftarrow 2$,
$\mathit{currIndexCount} \leftarrow 0$,
$\mathit{temp}_1$,
$\mathit{temp}_2$.

\nl {\bf Function $\mathit{Function\_delete}(T_i)$} \nllabel{ln:deleted}
\Begin{

\nl  \For{$ k\in \{1, iteration\}$ \nllabel{ln:outer_for_loop}}{


 \nl    \While{$ \mathit{currIndexCount} \leq n$ \nllabel{ln:outer_while_loop}}{


\nl    \While{$ \ell \in \{\mathit{currIndexCount}, \mathit{currIndexCount} + \mathit{blockSize}\}$ \nllabel{ln:inner_while_loop}}{

                \nl $\mathit{temp}_1 = T_i[r_{\ell}]$, $\mathit{temp}_2 = T_i[r_{\ell+stepSize}]$
				
				\nl $value \leftarrow \mathit{temp}_1$ \encircle{$\mathcal{H}$} $\mathit{temp}_2$
				
				\nl \nllabel{ln:array_store_1}$Temp\_array[\ell] \leftarrow value$

				\nl \nllabel{ln:array_store_2}$Temp\_array[\ell+\mathit{stepSize}] \leftarrow value$;

				\nl $\ell\leftarrow \ell+1$

				\nl \lIf{$(\ell+\mathit{stepSize} = \mathit{currIndexCount} + \mathit{blockSize})$}{ break}
        }
        \nl $\mathit{currIndexCount} \leftarrow \mathit{currIndexCount} + \mathit{blockSize}$
    }
    \nl \nllabel{ln:array_update}$T_i \leftarrow \mathit{Temp\_array}$

	\nl	\nllabel{ln:block_size_update} $\mathit{blockSize} \leftarrow \mathit{blockSize} \times 2$
	
	\nl \nllabel{ln:step_size_update}	$\mathit{stepSize} \leftarrow \mathit{stepSize} \times 2$
}
\nl $\mathit{Proof}_i \leftarrow \mathcal{H}(r_j)$, where $j\in \{1,n\}$ \nllabel{ln:compute_proof_delete}

\nl Write deleted sensor readings $r_j$ ($j\in \{1,n\}$) of the epoch $T_i$ and $\mathit{Proof}_i$ on the disk \nllabel{ln:write_data_on_disk_after_delete}
}
\caption{State transition phase.}
\label{alg:State_transition_ phase_execution}
\end{algorithm}
\setlength{\textfloatsep}{0pt}

\subsection{State Transition Phase}
\label{subsec:State Transition Phase}
In this phase, the sensor data belonging to an epoch is deleted, based on the data retention policy. Recall that (as mentioned in \S\ref{subsec:Scoping the Problem}) if data is replaced by null strings, then it can be recovered, as shown in~\cite{gutmann1996secure,DBLP:conf/fast/PetersonBHSR05}. Our proposed verifiable data deletion method (see Algorithm~\ref{alg:State_transition_ phase_execution}) guarantees the irrecoverability of the data by implementing a memory-hard function, provided by the SDP.

\medskip
\noindent\textit{\underline{Step 1: Selection of the epoch on which the deletion algorithm}} \textit{\underline{will be executed.}} As mentioned in \S\ref{subsec:Data Retention Policy}, all sensor readings belonging to an epoch $T_i$ are deleted by the cloud at the beginning of an epoch whose begin time is $T_i.\mathit{et} + \mathcal{P}_{\mathit{del}} \times \Delta$, where $T_i.\mathit{et}$ is the end time of the epoch $T_i$, $\mathcal{P}_{\mathit{del}}$ corresponds to the number of epochs after which the data must be deleted, and $\Delta$ is the duration of the epoch.

\medskip
\noindent\textit{\underline{Step 2: Deleting sensor reading (Lines~\ref{ln:deleted}-\ref{ln:step_size_update}).}} Suppose that in an epoch $T_i$, $n$ sensor records are needed to be deleted. The deletion function executes a cryptographic memory-hard function, denoted by $\encircle{$\mathcal{H}$}$\footnote{{\scriptsize For simplicity, we considered the one-way hash function, $\mathcal{H}$ for the construction of memory-hard function.}} on all the $n$ sensor readings in $\log n$ number of iterations and produces the final output in such a way that the original $n$ sensor readings cannot be obtained from the output.

\begin{figure}[!t]
\begin{center}
		\includegraphics[scale=0.4]{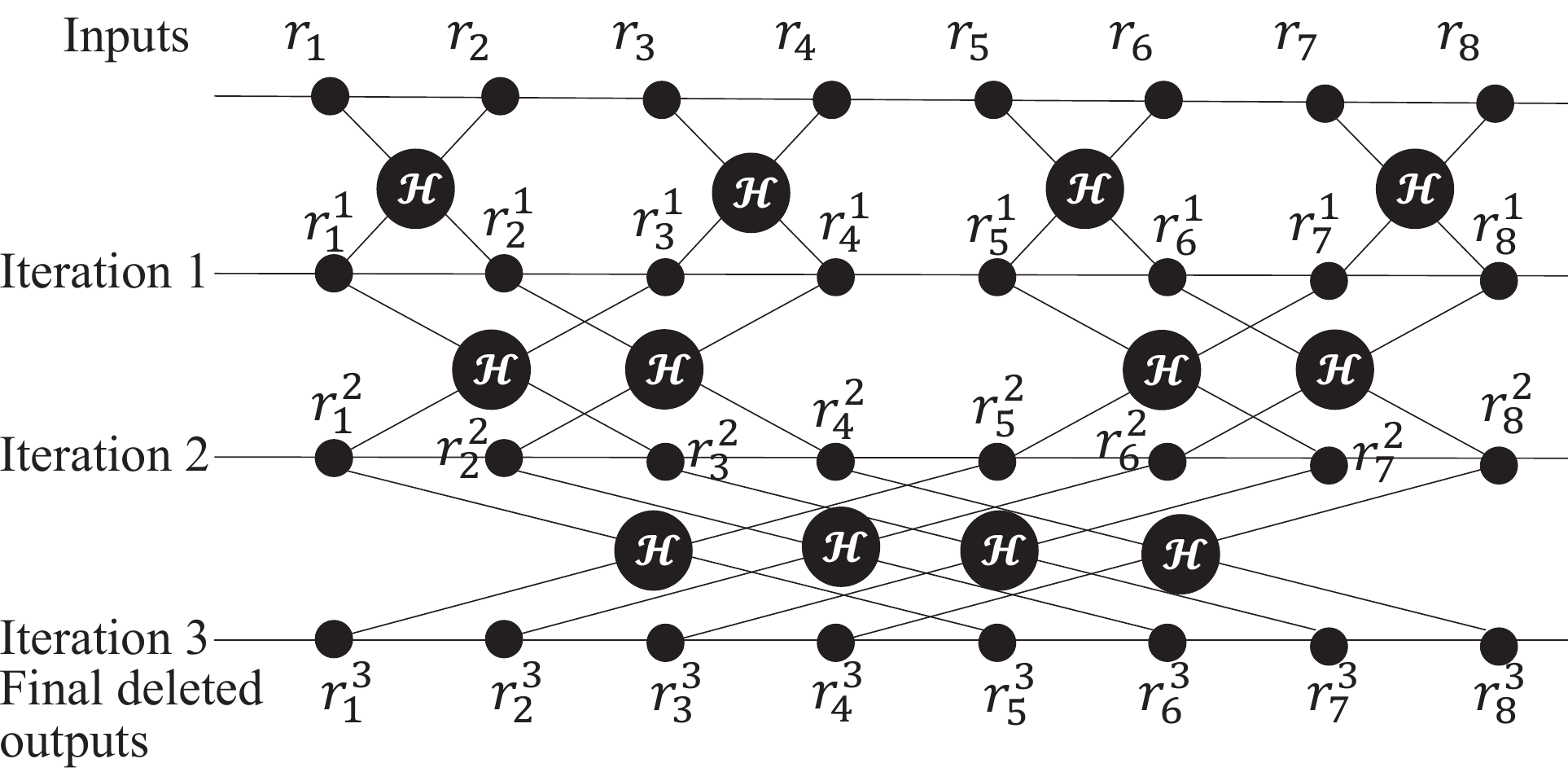}
	\end{center}
\BBB
	\caption{An illustration of delete algorithm execution.}
\label{fig:Delete algorithm execution}
\end{figure}

We explain the deletion steps with the help of an example, where an epoch $T_i$ contains eight sensor records that need to be deleted.
Algorithm~\ref{alg:State_transition_ phase_execution}
shows pseudocode of the deletion method, and the execution pattern of the algorithm is shown in Figure ~\ref{fig:Delete algorithm execution}. In Algorithm~\ref{alg:State_transition_ phase_execution}, Lines~\ref{ln:deleted}-\ref{ln:step_size_update} will be executed $\log n = \log 8 = 3$ times for the case of deleting eight sensor readings. The deletion algorithm divides the initial array of eight sensor readings into four blocks, where each block contains two sensor readings. Thus, the computation $\encircle{$\mathcal{H}$}$ is performed on each block having the following sensor records: $\langle r_1, r_2 \rangle$, $\langle r_3, r_4 \rangle$, $\langle r_5, r_6\rangle$, $\langle r_7, r_8 \rangle$ (according to the while-loop starting from Line~\ref{ln:inner_while_loop}). The newly computed sensor records are stored in a temporary array of length $n$ (as shown in Lines \ref{ln:array_store_1}, \ref{ln:array_store_2}). Thus, at the end of the first iteration, we obtain the following sensor records:
$\langle r_1^1,r_2^1\rangle$,
$\langle r_3^1,r_4^1\rangle$,
$\langle r_5^1,r_6^1\rangle$, and
$\langle r_7^1,r_8^1\rangle$.
At the end of each iterations, we increase the block size and the step size (\textit{i}.\textit{e}., a variable that is used to create a pair of sensor readings) by an order of two (see Lines~\ref{ln:block_size_update}-\ref{ln:step_size_update}). Further, at the end of each iteration, all the sensor readings of the epoch $T_i$ are overwritten by the newly computed sensor readings (see Line \ref{ln:array_update}).

In the second iteration, according to Line~\ref{ln:inner_while_loop}, the function is executed over the sensor records:
$\langle r_1^1,r_3^1\rangle$,
$\langle r_2^1,r_4^1\rangle$,
$\langle r_5^1,r_7^1\rangle$, and
$\langle r_6^1,r_8^1\rangle$, and produces the sensor records:
$\langle r_1^2,r_2^2\rangle$,
$\langle r_3^2,r_4^2\rangle$,
$\langle r_5^2,r_6^2\rangle$, and
$\langle r_7^2,r_8^2\rangle$.
In the third iteration, the function is executed over the sensor records: $\langle r_1^2,r_5^2\rangle$,
$\langle r_2^2,r_6^2\rangle$,
$\langle r_3^2,r_7^2\rangle$, and
$\langle r_4^2,r_8^2\rangle$, and produces the following sensor records as final output on which a hash function is executed to generate a proof of deletion:
$\langle r_1^3,r_2^3\rangle$,
$\langle r_3^3,r_4^3\rangle$,
$\langle r_5^3,r_6^3\rangle$, and
$\langle r_7^3,r_8^3\rangle$.

\medskip
\noindent\textit{\underline{Step 3: Computing a hash function to generate a proof of deletion}}  \textit{\underline{(Lines~\ref{ln:compute_proof_delete}-\ref{ln:write_data_on_disk_after_delete}).}} After executing step 2 on all the required sensor readings of the epoch $T_i$, the cloud executes a hash function on the outputs of step 2, and it produces a proof, $\mathit{Proof}_i$, of deletion. This proof is sent to the verifier during the attestation phase.

\medskip
\noindent\textbf{Note.} Unless the cloud executes the deletion method  (Algorithm~\ref{alg:State_transition_ phase_execution}), there is no value that can produce the proof of deletion that should also match with the verifiable tag, $\mathit{irH}$, which was already produced by the SDP. Also, note that our construction is based on memory-hard functions that require a significant amount of time to produce the proof in the above-mentioned steps 2 and 3, compared to transmitting the proof to the verifier in the attestation phase (see details in \S\ref{subsec:Attestation Phase}).

\subsection{Attestation Phase}
\label{subsec:Attestation Phase}
\textsc{IoT Expunge} allows users and the SDP to verify the data state against the \emph{pre-notified} data retention policy, as will be described in this section (see Algorithm~\ref{alg:verification algorithm}). First, we show that a user can verify the data state, and then, at the end of this section, also show how the SDP verifies the data state.

\medskip
\noindent\textbf{Verification of User-Associated Data.} The objectives of user-side verification are as follows: (\textit{i}) it needs to find the presence/absence of user-associated data in an epoch, and (\textit{ii}) it needs to verify the data state (accessible or irrecoverable) against the data retention policies. Below, we discuss two cases, when the user wishes to verify her data and the state of the data.

\DontPrintSemicolon
\LinesNotNumbered
\begin{algorithm}[!t]
\textbf{Inputs:} User device $d_u$, Epoch $T_j$, hash digests $h^y_j$,
$\mathcal{E}_{\mathit{PK}_{E}}(r_y)$,
$1\leq y\leq n$,
$\mathcal{CT}_j$,
$\mathcal{CT}_{j-1}$,
$\mathcal{E}_{\mathcal{K}}(\mathcal{CT}_j)$,
$\mathcal{E}_{\mathcal{K}}(\mathit{aH}_j)$ \\
Hash function: $\mathcal{H}$. Public values: $x$ and $\eta$. $\mathit{Decrypt}()$: A decryption function
	
\nl{\bf Function $\mathit{Verify}(\mathcal{E}_{PK_{E}}(r_y))$} \nllabel{ln:verify}
\Begin{

\nl \For{$y\in \{1,n\}$ \nllabel{ln:checking_record_presence}}{
		
\nl $h^y \leftarrow \mathcal{H}(d_u,T_j,y)$

\nl \lIf{$ h^y = h^y_j$}{
The user-associated data exists in the epoch $T_j$
\nllabel{ln:user-associated_data_exist}}

\nl \lElse{
The user-associated data does not exist in the epoch $T_j$
\nllabel{ln:user-associated_data_NOT_exist}}}

\nl \lIf{$T_j \wedge j=1$}{
$\alpha\leftarrow[x^{\mathcal{H}(h^1||h^2||\ldots||h^n)}]\bmod \eta$
\nllabel{ln:alpha_for_FIRST_epoch}}

\nl \lElse{ $\alpha\leftarrow[\mathcal{CT}_{j-1}^{\mathcal{H}(h^1||h^2||\ldots||h^n)}]\bmod \eta$
\nllabel{ln:alpha_for_OTHER_epoch}}

\nl \lIf{$\alpha=\mathcal{CT}_j$}{
The verifier has received all sensor records belonging to the desired epoch $T_j$ \nllabel{ln:verifier_got_all_records}}

\nl $\mathit{aH}_j \leftarrow Decrypt[\mathcal{E}_{\mathcal{K}}(\mathit{aH}_j)]$ \nllabel{ln:decrypt_ah}

\nl $\mathit{uH}\leftarrow\mathcal{H}[\mathcal{E}_{\mathit{PK}_E}(r_y)]: y\in\{1,n\}$ \nllabel{ln:compute_uH}
		
\nl \If{$\mathit{uH}=\mathit{aH}_j=$ Data retention policy \nllabel{ln:verify_policy}}{
Sensor readings are in accessible state, and the cloud is keeping the data against the data retention policy }
		}
\caption{Attestation phase.}
\label{alg:verification algorithm}
\end{algorithm}
\setlength{\textfloatsep}{0pt}

\medskip
\noindent\textbf{\textsc{Verification in the accessible state.}} We first consider the case of verifying the data that is in accessible state. Let $t_i$ be the time for which the user wishes to verify his/her records. The user executes the following steps:

\smallskip
\noindent\textit{\underline{Step 1: Request the cloud to send data.}} In this step, the user specifies the desired timestamp $t_i$ to the cloud. In response, the cloud sends the following data from the relation \texttt{SensorData}: (\textit{i}) all the encrypted sensor records that belong to an epoch, say $T_j$, that contains the requested sensor reading having time $t_i$, (\textit{ii}) the epoch-id, say $T_j$, (\textit{iii}) cryptographic timestamp, say $\mathcal{CT}_j$, (\textit{iv}) the hash digest $h_j^y$ (where $1\leq y\leq n$, $n$ is the number of sensor readings) of sensor readings of the epoch $T_j$,
(\textit{v}) the cryptographic time, say $\mathcal{CT}_{j-1}$, of the previous epoch, say $T_{j-1}$, if exists, and
(\textit{vi}) from the relation \texttt{MetaData}: epoch-id $T_j$, encrypted cryptographic timestamp $\mathcal{E}(\mathcal{CT}_j)$, and the encrypted verifiable tag, say $\mathcal{E}(\mathit{aH}_j)$.

\smallskip
\noindent\textit{\underline{Step 2: Verification of presence/absence of user-associated data.}}  \textit{\underline{(Lines~\ref{ln:checking_record_presence}-\ref{ln:user-associated_data_NOT_exist}).}}
In this step, the user verifies the presence/absence of her data in the encrypted sensor records at the cloud. The user knows her device-id, say $d_u$, and hence, the user executes the hash function, $\mathcal{H}$, to know the presence/absence of her sensor data, as follows:

\centerline{
$h^1 \leftarrow \mathcal{H}(d_u,T_j,1)$, \\
$h^2 \leftarrow \mathcal{H}(d_u,T_j,2)$, \\
$\ldots$, \\
$h^n \leftarrow \mathcal{H}(d_u,T_j,n)$ \\
}
Where $T_j$ is the epoch-id received from the cloud, and $n$ is the number of encrypted sensor readings in the epoch received from the cloud. The user matches each computed hash digest $h^y$ against $h^y_j$, where $1\leq y\leq n$. If any two hash digests match, it shows that the user-associated data is present in some of $n$ sensor readings.


\smallskip
\noindent\textit{\underline{Step 3: Verification of the completeness of received sensor readings.}}  \textit{\underline{(Lines~\ref{ln:alpha_for_FIRST_epoch}-\ref{ln:verifier_got_all_records}).}} Proving the presence/absence of the user-associated data does not prove that the user has received all the sensor readings of epoch $T_j$ (requested by the user). Thus, the user also verifies the completeness of sensor readings from the epoch (\textit{i}.\textit{e}., the user has received all the encrypted sensor readings belonging to the epoch $T_j$), as follows:

\centerline{
$\alpha \leftarrow [\mathcal{CT}_{i-1}^{\mathcal{H}(h^1||h^2||\ldots ||h^n)} ]\bmod \eta$}
The user compares $\alpha$ against the cryptographic time $\mathcal{CT}_j$, and if they match, it shows the user has received all sensor readings of the desired epoch (Line~\ref{ln:verifier_got_all_records}).

\smallskip
\noindent\textit{\underline{Step 4: Verification of data state (Lines~\ref{ln:decrypt_ah}-\ref{ln:verify_policy}).}} Finally, the user wishes to verify the data state against the pre-notified data retention policies. The user executes the hash function on the received encrypted sensor readings, and matches the computed hash digest, say $\mathit{uH}$, against the decrypted value of $\mathcal{E}(\mathit{aH}_j)$, denoted by $\mathit{aH}_j$. If both the hash digests match, it shows that the data state is accessible.

\medskip
\noindent\textbf{Information leakage discussion.}
Recall that a verifier receives the sensor readings in the encrypted format; hence, the verifier cannot learn the cleartext sensor data. Further, based on the received hash digests from the cloud, the verifier cannot learn how many other users associated data are present in the data. The reason is: each hash digest is different, and the verifier is not aware of other users' device-ids. Thus, our verification method does not reveal any information to the verifier about other users.

\medskip
\textbf{Aside.} A similar method can also be executed for verifying user-associated data in a time range.

\medskip
\noindent\textbf{\textsc{Verification in the irrecoverable state.}} We next discuss how the verifier can attest that the cloud has deleted the data based on the data retention policy. Here, the objective of the verification is almost the same (as in the previous case of verifying accessible state data), \textit{i}.\textit{e}., verifying the existence of user-associated data in an epoch and verifying the data state to be irrecoverable. Thus, in this case, steps 1, 2, and 3 are executed like the previous case of verifying accessible state of the data. However, in step 1, the cloud will also send the encrypted verifiable tag $\mathcal{E}(\mathit{irH}_i)$ (from the \texttt{MetaData} relation) of the desired epoch, say $T_i$, (instead of the encrypted verifiable tag $\mathcal{E}(\mathit{aH}_i)$).

\smallskip
\noindent\textit{\underline{Step 4: Verification of deletion.}} In this step, the user verifies the time-bounded response from the cloud to deduce that the cloud has deleted the data by following the data retention policy, not when the verification request is arriving from the user. The time-bounded delay in proof generation, \textit{i}.\textit{e}., the hash digest over all the deleted rows (here denoted by $\mathit{Proof}_i$) for the desired epoch $T_i$ (refer to steps 2 and 3 in \S\ref{subsec:State Transition Phase}), at the cloud, identifies the possibility whether the cloud is generating the proof ($\mathit{Proof}_i$) on-the-fly after receiving the verification request from the user or the cloud has already generated the proof ($\mathit{Proof}_i$) by deleting the data against the data retention policy.

Note that in the case when the cloud has not deleted the sensor readings of the desired epoch, the cloud will compute the proof ($\mathit{Proof}_i$) by executing the deletion algorithm (\textit{i}.\textit{e}., memory-hard functions). However, the computation of the deletion algorithm and generation of the proof ($\mathit{Proof}_i$) will take a longer time compared to transmitting the already computed proof (as mentioned in step 3 in \S\ref{subsec:State Transition Phase}). Further, the cloud also sends the encrypted verifiable tag $\mathcal{E}(\mathit{irH}_i)$ (which, recall that, was outsourced by the SDP in \textsc{Stage} 3 of \S\ref{subsec:Control Phase}) from the \texttt{MetaData} relation. In this step, the user matches the proof of deletion $\mathit{Proof}_i$ with the decrypted value of $\mathcal{E}(\mathit{irH}_i)$. If both the value matches, then it shows that the cloud has deleted the data against the data retention policy.

\smallskip
\noindent\textbf{Verification by the SDP.} Our approach also allows the SDP to verify the data state against the data retention policies by executing steps 1, 3, and 4. Note that the SDP does not need to execute step 2.

\DontPrintSemicolon
\LinesNotNumbered
\begin{algorithm}[!t]
\textbf{Inputs:} Block: $B_i$. Query records: $\langle q_j,t_j,u_j \rangle$, where $j\in \{1,n\}$

\nl{\bf Function $\mathit{QueryLog}(B_i)$} \nllabel{ln:qlog}
\Begin{

\nl \For{$j\in \{1,n\}$ \nllabel{ln:doing_hash_chain}}{

\nl \lIf{$j\neq1$}{
$Bh^j_i\leftarrow \mathcal{H}(q_j,t_j,u_j||Bh^{j-1}_i)$
\nllabel{ln:hash_chain_for_non1}}

\nl \lElse{$Bh^j_i\leftarrow \mathcal{H}(q_j,t_j,u_j||x)$
\nllabel{ln:hash_chain_for_1}}}

\nl \lIf{$T_i\wedge (i=1)$ \nllabel{ln:proof_creation}}{
$\mathit{BProof}_i\leftarrow[x^{Bh^n_i}]\bmod \eta$ \nllabel{ln:proof_for_1}}

\nl \lElse{$\mathit{BProof}_i\leftarrow[\mathit{BProof^{Bh^n_i}_{i-1}}]\bmod \eta$ \nllabel{ln:proof_for_non1}}

\nl Write encrypted block $B_i$ and $\mathit{BProof}_i$ on disk \nllabel{ln:writing_to_disk}
}
\caption{Query logging phase.}
\label{alg:Query_logging_phase}
\end{algorithm}
\setlength{\textfloatsep}{0pt}

\subsection{Query Logging Phase}
\label{subsec:Query Log Phase}
This section provides a method (see Algorithm~\ref{alg:Query_logging_phase}) for securely storing all incoming
queries to produce tamper-proof query logs and a method to verify the query logs by the SDP. Recall that the reason of having and verifying query logs is to know whether the queries are requested by the user or the SP is executing the query to learn the sensor data. In short, the query logging phase includes the following stages:
creating a block of queries (\textsc{Stage} 1),
creating a hash chain over the queries in a block (\textsc{Stage} 2), and generating a block proof for each block (\textsc{Stage} 3). As it will be clear soon that the purpose of creating hash chains and block proofs is to detect that the SP is not deleting any query belonging to the block, as well as, not deleting any block. Below, we explain all three stages:

\medskip
\noindent\textbf{\textsc{Stage 1}: Block selection.} Since we cannot store all queries from the users inside the enclave due to its limited memory, we need to write the queries in a secure and tamper-proof manner on disk, which is managed by the SP that can tamper with the queries. However, creating secure logs having all queries incurs the overhead on the verifier. Thus, we need to select a fixed-size memory block that should be less than the enclave memory. The block is used to store the queries inside the enclave and in encrypted form on the disk. We denote an $i^{\mathit{th}}$ block by $B_i$. Each block contains its creation time, using which the verifier can verify a particular block for the desired time. An entry in a block is a query record, denoted by $\langle q_i, t_i, u_j \rangle$, where $q_i$ is the $i^{\mathit{th}}$ query arrived from the user $u_j$ at time $t_i$. Particularly, $u_j$ indicates proof of identity of the user $u_j$, thereby the user $u_j$ cannot deny later after transmitting the query to the SP.

The block size may depend on several factors, \textit{e}.\textit{g}., the time duration, the enclave size, the verification time for verifying the block, and the communication cost for moving the block during verification from the SP to the verifier. A small-sized block minimizes the above-mentioned last two factors, by avoiding verifying the entire query log, which may span over many years in a practical system.

\medskip
\noindent\textbf{\textsc{Stage 2}: Hash-chain creation (Lines~\ref{ln:doing_hash_chain}-\ref{ln:hash_chain_for_1} of Algorithm~\ref{alg:Query_logging_phase}).} This stage works in a similar way as \textsc{Stage} 2 in \S\ref{subsec:Control Phase}.

\smallskip
\noindent\textit{\underline{Step 1: Initialization.}} This step is identical to step 1, as in \S\ref{subsec:Attestation Phase} to generate a seed value $x$ using a PRG function and two large prime numbers $p$ and $q$, such that $\eta=p\times q$. 

\smallskip
\noindent\textit{\underline{Step 2: Hash chain creation.}} This step creates a hash chain over all the query records in a block. Consider that $n$ query records exist in the first block $B_1$. The enclave creates a hash chain over query records, as follows:

\centerline{
$Bh^1_1 \leftarrow \mathcal{H}(q_1, t_1, u_1 || x)$,}
\centerline{
$Bh^2_1 \leftarrow \mathcal{H}(q_2, t_2, u_2 || Bh^1_1 )$,}
\centerline{$\vdots$}
\centerline{$Bh^n_1 \leftarrow \mathcal{H}(q_2, t_2, u_2 || Bh^n_1)$ }
Where $Bh_i^j$ denotes the hash digest for the $j^{\mathit{th}}$ query record in the $i^{\mathit{th}}$ block. Note that the hash digest of the $i^{\mathit{th}}$ query record is taken with the
$(i+1)^{\mathit{th}}$ query record, when computing a hash digest for the query record $i+1$, except for the first query record, where we used the random number $x$.

\medskip
\noindent\textbf{\textsc{Stage 3}: Block proof creation (Lines~\ref{ln:proof_for_1}-\ref{ln:proof_for_non1} of Algorithm~\ref{alg:Query_logging_phase}).} For a block $B_i$, after computing the hash digest for the last query record, we compute a proof for the block $B_i$. Here, we show how the enclave creates the proof for the first block $B_1$ and the second block $B_2$. A similar method is used over other blocks too. Let $Bh^n_1$ and $Bh^n_2$ be the hash digests computed for the last $n^{\mathit{th}}$ query records of the blocks $B_1$ and $B_2$. Note that for simplicity, we assumed that the block contains an identical number of query records. For the block $B_1$, the proof (denoted by $\mathit{BProof}_1$) is created as follows (Line~\ref{ln:proof_for_1}):

\centerline{$\mathit{BProof}_1 \leftarrow [x^{Bh^n_1}]\bmod \eta$}
Now, to generate the proof for the second block $B_2$, we use the proof of the previous block, \textit{i}.\textit{e}., $\mathit{BProof}_1$, and it creates a chain over the block proofs, as follows (Line~\ref{ln:proof_for_non1}):

\centerline{$\mathit{BProof}_2 \leftarrow [\mathit{BProof}_1 ^{Bh^n_2}]\bmod \eta$}

\medskip
\noindent\textbf{\textsc{Stage} 4: Writing data to disk (Line~\ref{ln:writing_to_disk} of Algorithm~\ref{alg:Query_logging_phase}).} The enclave writes the following on the disk:
(\textit{i}) a block $B_i$, $i>0$, having query records encrypted using the public key of the SDP (denoted by $\mathcal{E}_{{\mathit{PK}}_{\mathit{SDP}}}(q_i, t_i, u_j)$), and
(\textit{ii}) the block proof $\mathit{BProof}_i$.

\begin{figure*}[!t]
		\BBB\BBB\B
		\begin{center}
			\begin{minipage}{.24\linewidth}
				\centering
	    \includegraphics[scale=0.26]{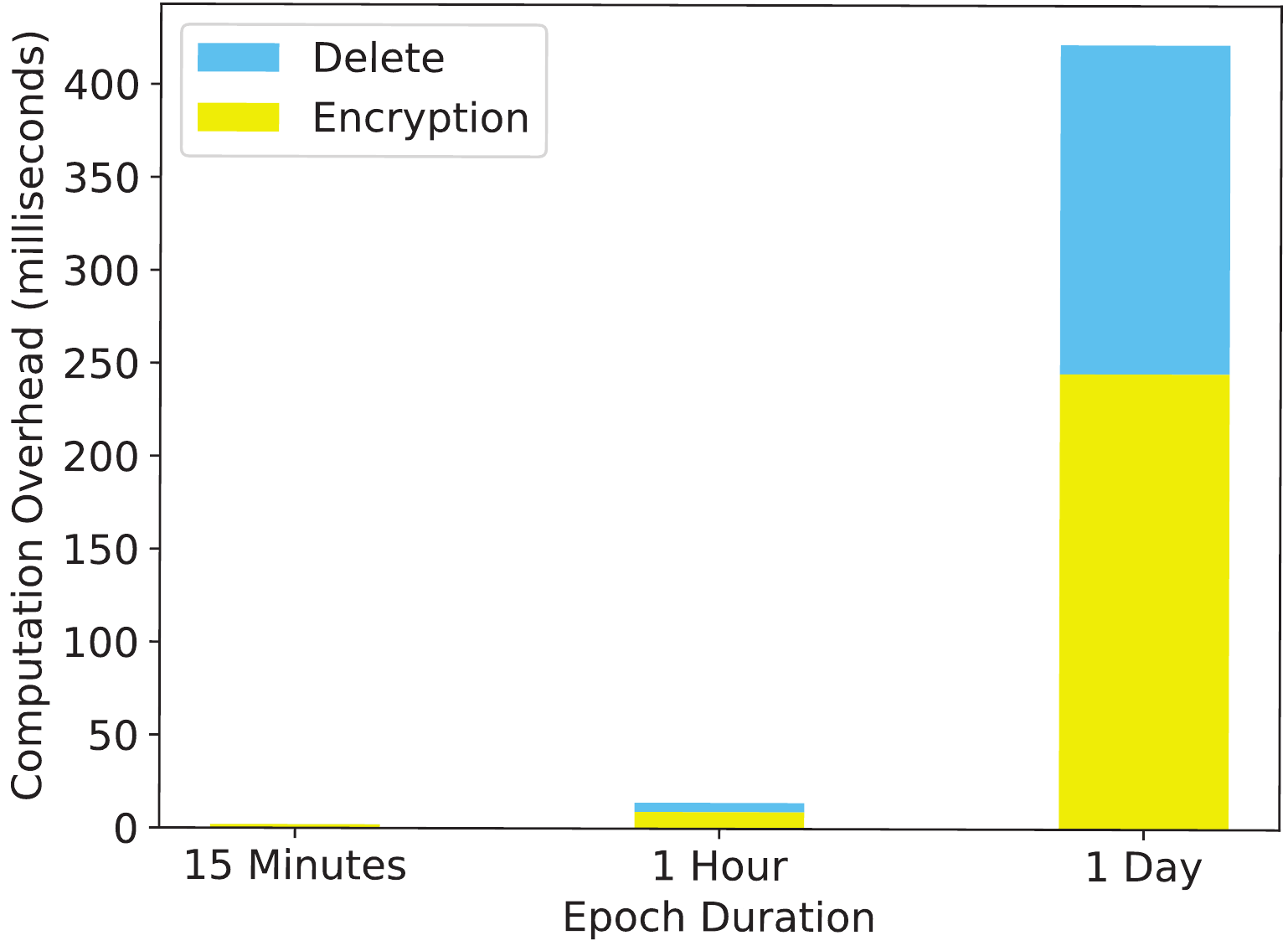}
       \BB\BBB\B
        \caption{Computation time at the SDP: Overhead per epoch.}
        \label{fig:Computational_overhead_SDP_perblock}
			\end{minipage}
			\begin{minipage}{.24\linewidth}
				\centering
	    \includegraphics[scale=0.26]{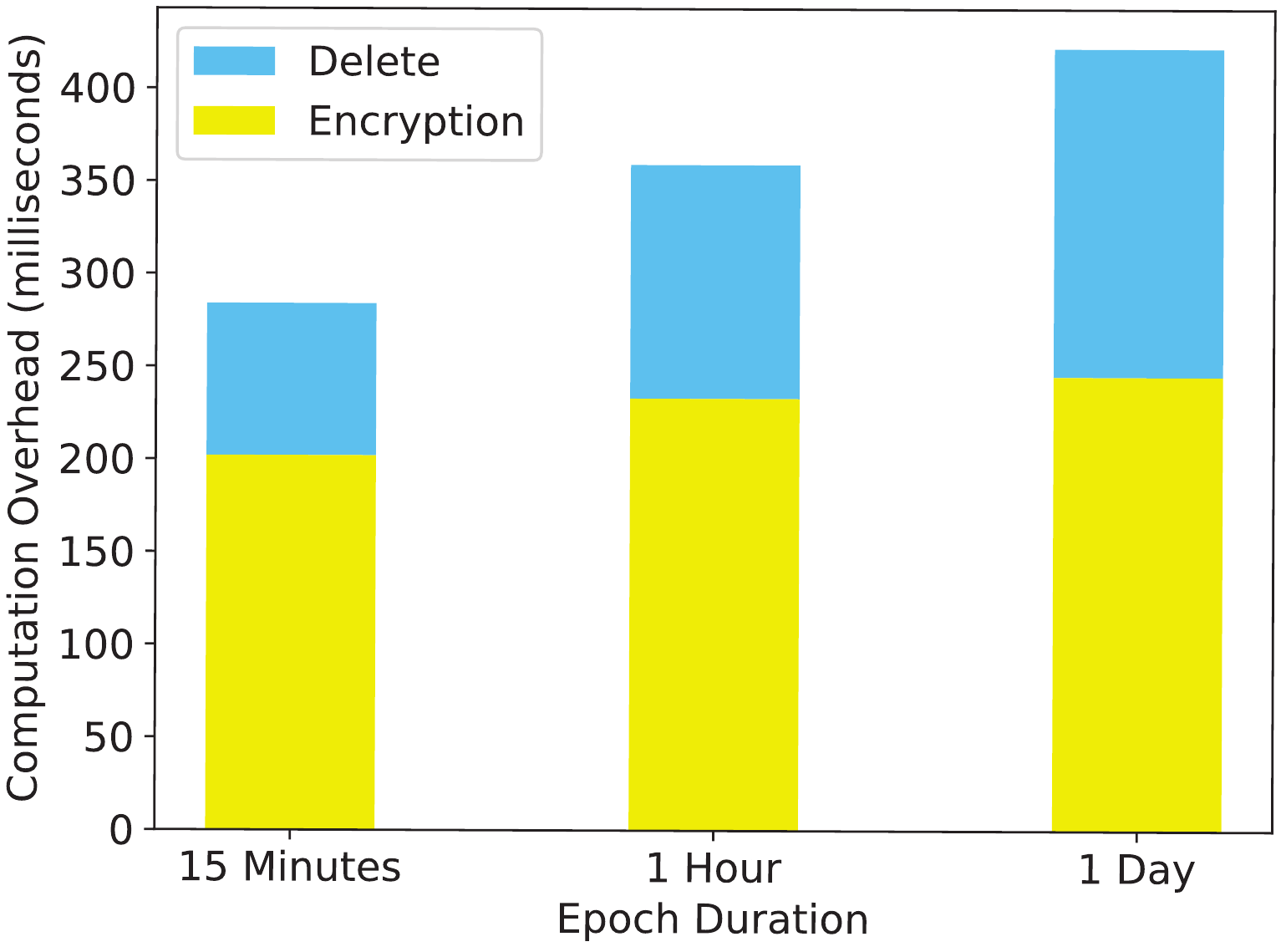}
       \BB\BBB\B
        \caption{Computation time at the SDP: Overhead per day.}
        \label{fig:Computational_overhead_SDP_perday}
			\end{minipage}			
			\begin{minipage}{.24\linewidth}
				\centering
            	\includegraphics[scale=0.26]{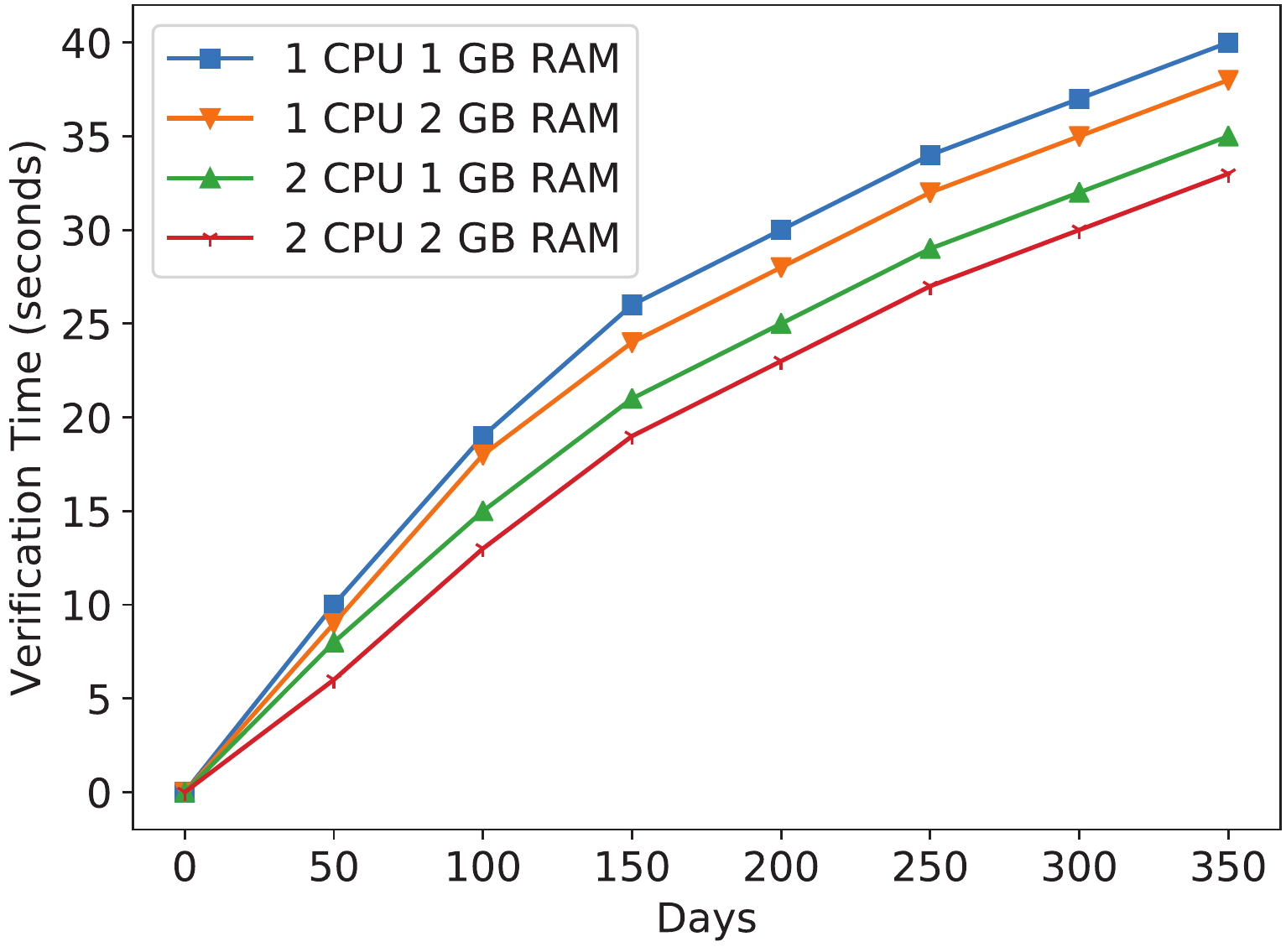}
               \BB\BBB\B
                \caption{Verification time at different users.}
                \label{fig:User verification time}
			\end{minipage}
			\begin{minipage}{.24\linewidth}
				\centering
            	\includegraphics[scale=0.26]{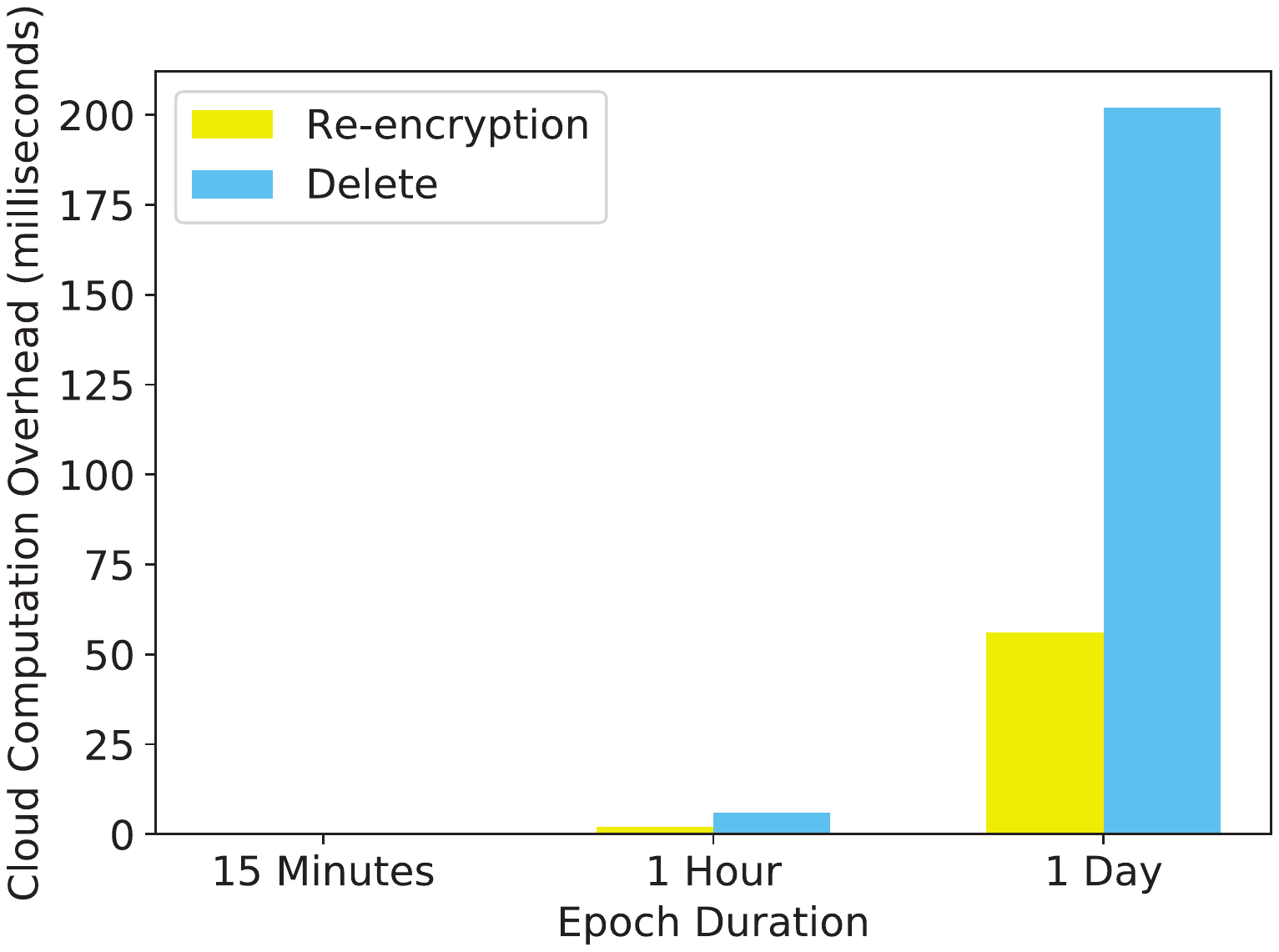}
            \BB\BBB\B
                \caption{Computational overhead at the cloud -- per epoch overhead.}
                \label{fig:cloud_overhead_per_block}
            \end{minipage}
		\end{center}
		\BBB\BBB
	\end{figure*}

\smallskip
\noindent
\textbf{Note: Verification of query log by the SDP.} The SDP requests the SP to send the following: (\textit{i}) encrypted query records of the desired block, say $B_i$, (\textit{ii}) the block proof of the block $B_i$, \textit{i}.\textit{e}., $\mathit{BProof}_i$, and (\textit{iii}) the block proof of the previous block $B_{i-1}$, \textit{i}.\textit{e}., $\mathit{BProof}_{i-1}$. On receiving the query records, the SDP decrypts them. On decrypting, the SDP may check whether the user has executed the query or the SP. Further, to ensure that the arrived query records are correct and complete, the SDP executes the above-mentioned step 2 of \textsc{Stage} 2 and \textsc{Stage} 3. It results in a proof, say $\mathit{Proof}_{\mathit{SDP}}$. The SDP matches $\mathit{Proof}_{\mathit{SDP}}$ with $\mathit{BProof}_i$, and if they match, it results in that the SP has not tampered with any query record.

\medskip
\noindent\textbf{Note: Dealing with multiple service providers.} In the full version~\cite{TR2020}, we show how to extend \textsc{IoT Expunge}
for multiple SPs having different data retention policies.

\section{Experimental Evaluation}
\label{sec:Experimental Evaluation}

We conducted an experimental evaluation of \textsc{IoT Expunge} over our campus testbed, which we alluded to in the introduction. To provide context, we first discuss the university testbed and then describe our experiments.

\subsection{TIPPERS System}
\label{subsec:Testbed Description}
TIPPERS System is a smart space middleware that provides campus-level location-based services (both inside and outside buildings) using WiFi access-point connectivity data. In our university, the Office of Information Technology (OIT) manages more than 2000 WiFi access-points that are connected to four WLAN controllers to provide campus-wide wireless network coverage in the campus. When a device gets connected to the university WiFi network (through an access-point $s_i$), the access-point $s_i$ generates Simple Network Management Protocol (SNMP) trap for this association event that produces a tuple of the form $\langle s_i,d_j,t_k\rangle$, where $d_j$ is the user device MAC address that is connected to the access-point $s_i$ at time $t_k$. In real-time, all SNMP traps $\langle s_i,d_j,t_k\rangle$ are sent to the access-point's controller that forwards such traps (after anonymizing the device id) to the forwarding server located at OIT. This WiFi connectivity data is sent to research groups or service providers. One such a research group (or a service provider) is a campus-level smart system, we have built, \textit{i}.\textit{e}., TIPPERS, which uses WiFi connectivity data to build applications, such as real-time occupancy of different regions/buildings, longitudinal analysis of building occupancy, and live heat map at the university campus scale.

The campus administration, through its privacy and security committee, imposed a key requirement on OIT that it must ensure that outsourced encrypted WiFi data is deleted from the storage based on the retention policy. \textsc{IoT Expunge} was motivated by the above requirement. In addition to implementing retention policies, we also developed mechanisms to ensure that all data access at the service provider (viz. TIPPERS system) are logged in a tamper-proof manner with verifiable proofs of access. Such a mechanism can be used to verify that the requested services/queries are generated by the user, and the service provider is not executing the services on its own to learn the behavior of WiFi users
in the campus. The implementation of the retention policy, coupled with the verification of access by SP, provides a secure solution that realizes (and goes beyond) the campus's data-sharing requirements.

\subsection{Experimental Results}
\label{subsec:Experimental Results}
To experiment with \textsc{IoT Expunge}, we worked with OIT, wherein OIT played the role of an SDP. OIT uses a timestamping server with 4 cores and 16 GB RAM. We used SHA-128 as the hashing algorithm.\footnote{One can use a different hashing algorithm too.} OIT distributed the desired security keys to the desired entities, as mentioned in \S\ref{sec:complete protocol}. After executing the control phase, the desired encrypted data (as mentioned in \S\ref{subsec:Control Phase}) is outsourced to a cloud machine of 8 cores and 32 GB RAM. The cloud forwards the encrypted sensor data to TIPPERS (or an SP) in a real-time manner.

\noindent\textbf{Dataset size and data retention policies.} Although \textsc{IoT Expunge} considers streaming WiFi data, in this section, we will provide the experimental results using the data collected over the past 12 months. The size of the original data was 1.8GB. The selected epoch durations are as follows: 15-minutes, 1-hour, and 1-day. The data retention policy, for the TIPPERS system, was set to keep only the last two days data in accessible state, and all the remaining data was expunged. 
Thus, for example, in the case of 1-hour epoch duration, the retention policy specifies that the data can be deleted after the arrival of next 48 epochs, each of duration 1-hour, while in the case of 1-day epoch duration, the retention policy specifies that data needs to be deleted after the arrival of two epochs, each of 1-day duration. The verification part of the retention policy was kept as infinity, since we primarily focused on testing performance of verification and logging. The verification part of the retention policy only affects storage at the cloud.

\medskip
\noindent\textbf{Exp 1. Computational time at the SDP.} Figures~\ref{fig:Computational_overhead_SDP_perblock} and~\ref{fig:Computational_overhead_SDP_perday} show the computational time at the SDP for executing the control phase. First, we measure the computational time for encrypting the sensor readings and generating the verifiable tag for deletion, for each epoch of duration 15-min, 1-hour, and 1-day; see Figure~\ref{fig:Computational_overhead_SDP_perblock}. Then, we measure the computational time for encrypting the sensor readings and generating the verifiable tag for 1-day data using epochs of different durations (15-min, 1-hour, and 1-day); see Figure~\ref{fig:Computational_overhead_SDP_perday}. Figure~\ref{fig:Computational_overhead_SDP_perblock} shows that as the epoch size increases, the computational time also increases, since each epoch contains more sensor readings. Figure~\ref{fig:Computational_overhead_SDP_perday} shows that having encrypted data for 1-day either in the form of epochs of 15-min, 1-hour, or 1-day, takes almost a similar time in encryption, while encrypting different number of epochs, such as 96 epochs (in case of 15-min epoch duration), 24 epochs (in case of 1-hour epoch duration), and 1 epoch (in case of 1-day epoch duration). However, for 1-day time period, using epochs of 15-min, 1-hour, and 1-day durations, generating verifiable tags takes a different amount of time. The reason is: in the case of 96 epochs, each of 15-min duration, we compute the verifiable tag for deletion on fewer number of sensor readings of each epoch, compared to having only 1 epoch for the entire day. Figure~\ref{fig:Computational_overhead_SDP_perday} shows that \emph{as the size of the epoch increases, generating tags for a fixed duration also increases}.

\medskip
\noindent\textbf{Exp 2. Storage requirement at the cloud.} The storage at the cloud in the case of different epochs having different durations (15-min, 1-hour, and 1-day) was almost identical.\footnote{For one of verifiable-tags, our dataset produced 360, 8640, 34500 verifiable-tags for 1-day, 1-hour, 15-min epochs, and took 6KB, 139KB, 552KB, respectively.} We outsourced data of around 1 year. The \emph{raw} data, \textit{i}.\textit{e.}, the original sensor data without executing \textsc{IoT Expunge} over them took around 1.8GB. However, after executing the control phase of \textsc{IoT Expunge}, the data size increased to 2.4GB. It shows that the proposed approach does not require more storage space to keep hash digests and verifiable tags.

\medskip
\noindent\textbf{Exp 3. Verification at a resource-constrained user.} We considered different resource-constrained users to realize the practicality of \textsc{IoT Expunge}. Particularly, we considered four types of users based on different computational capabilities (\textit{e}.\textit{g}., available main memory -- 1GB/2GB, and the number of cores -- 1/2). Figure~\ref{fig:User verification time} shows the time of verification when data is in accessible state. Note that verifying 1-day data at resource-constrained users took at most 2seconds. As the number of days increases, the verification time also increases. Also, verifying 1-year data took less than 1-minute. Note that here we are not showing the time of verifying the deleted data, since it is based on a time-bounded response by the cloud, in which case the communication cost was negligible, \textit{i}.\textit{e}., 0.0007seconds (Exp 5), compared to producing the proof at the cloud in 1 second (Exp 4). However, in the case of 1-year data, the time of generating the deletion proof at the cloud took around 1-min, while transferring the deletion proof took only 0.0007seconds.

\medskip
\noindent\textbf{Exp 4. Performance at the cloud.} In \textsc{IoT Expunge}, the cloud executed: deletion operation on the data (and re-encryption of the data). As expected, the cloud took less time to execute both the operations on the small-sized epoch, compared to a larger-sized epoch, since the small-sized epoch stores less number of rows, and hence, the operation is executed on less number of rows compared to a larger-sized epoch; see Figure~\ref{fig:cloud_overhead_per_block}.

\medskip
\noindent\textbf{Exp 5. Communication overhead during verification.} We measured the communication impact when a verifier downloaded the sensor data. Consider a case when the verifier wishes to attest only one-hour/one-day data. The average size of one-hour (one-day) data was 0.2MB (5MB). When using slow (100MB/s), medium (500MB/s), and fast (1GB/s) speed of data transmission, the data transmission time in case of 1-hour or 1-day data was negligible.

\section{Related Work}
\label{sec:Related Work}
There has been tremendous research on IoT data processing and secure access~\cite{pill,tap} and privacy-preserving access to comply with the General Data Protection Regulation (GDPR)~\cite{primapotion,DBLP:conf/fc/BasinDH18}. A variety of data deletion methods have been proposed such that hardware sabotage, recoverable deletion through unlinking and re-encryption~\cite{rear,basin,DBLP:conf/fast/PetersonBHSR05,radia}. The survey in~\cite{surv} presents an overview of the existing techniques for secure deletion.

\noindent\textit{Encryption- or secret-sharing-based deletion.} The na\"{\i}ve solution to encrypt the data and then erasing the encryption key to render the ciphered data, is useless, since the problem can be reduced into recovering the erased secret key, which makes the data recoverable. 
Neuralyzer~\cite{lyzer} guarantees data deletion based on revoking access to the decryption key. Thus, the decryption key can be distributed among multiple peers of secret-shared form~\cite{DBLP:journals/cacm/Shamir79} to avoid key re-construction, unless peers collude with each other, which is hard to guarantee. Another solution is to store the secret key at the trusted platform module (TPM) and, then, guaranteeing the deletion operation execution inside TPM. For example,~\cite{pruf,delveri} proposed deletion through proof-of-work that enables a user to verify the correct implementation of cryptographic operations inside TPM, without having to access its internal source code. Speed~\cite{ammar} provides the trusted memory- and distance-bounding-based deletion.~\cite{repo,basin} has shown irrecoverable deletion through overwriting the storage media. Similarly,~\cite{ammar,sudik} have proposed deletion through overwriting over small capacity embedded devices. 

\noindent\textit{Blockchain-based deletion.}~\cite{block} proposed secure deletion using blockchains such that each deletion transaction is logged on the shared ledger that can be verified later. Also, recently, an integrated timestamping approach~\cite{bloo} has been brought into light that suggests combining the trustworthiness of the central solution with the scalability of de-centralized solutions. In particular, a blockchain-based timestamping solution can leverage the role of SDP by maintaining a public ledger of timestamps, where every time the SDP generates a unique timestamp, it must be published in the subsequent block of the public ledger. Therefore, these timestamps can be verified whenever the corresponding block is published on the main chain. The work in~\cite{nikol} presents a graph pebbling technique to ensure the data erasure in a bounded-space.

\noindent\textit{Caching vs retention.} A data retention policy can be considered as the proof of data possession over a function of time, and data retention policies are substantially different from the well-known caching policies~\cite{hash}. In general, caching satisfies future data requests to improve the performance by limiting the disk access and can be viewed as short-term dynamic data retention that does not consider the privacy aspect over past data.

\section{Conclusion}
\label{sec:conclusion}
We presented a framework, \textsc{IoT Expunge} for IoT data storage at the cloud against data retention policies. By implementing data retention policies, the data changes its state from accessible to irrecoverable, \textit{i}.\textit{e}., secure deletion. We provide a verifiable deletion method that can be executed at any third party without revealing data privacy. We have tested \textsc{IoT Expunge} in a real university-based smart space project, namely the TIPPERS system. The nominal verification time shows the practicality of \textsc{IoT Expunge}.

\begin{spacing}{0.83}
{\scriptsize

}
\end{spacing}

\end{document}